\documentclass[fleqn,usenatbib]{mnras}

\usepackage{newtxtext,newtxmath}
\usepackage{float}

\usepackage[T1]{fontenc}

\DeclareRobustCommand{\VAN}[3]{#2}
\let\VANthebibliography\thebibliography
\def\thebibliography{\DeclareRobustCommand{\VAN}[3]{##3}\VANthebibliography}

\usepackage{graphicx}	
\usepackage{amsmath}	
\title[The proto-galaxy of Milky Way-mass haloes]{The proto-galaxy of Milky Way-mass haloes in the FIRE simulations}

\author[D. Horta et al.]{
Danny Horta$^{1}$\thanks{E-mail: dhortadarrington@flatironinstitute.org}, 
Emily C. Cunningham$^{1,2 \dagger}$,
Robyn Sanderson$^{3}$,
Kathryn V. Johnston$^{2}$,
Alis Deason$^{4,5}$,
\newauthor
Andrew Wetzel$^{6}$,
Fiona McCluskey$^{6}$,
Nicol\'as Garavito-Camargo$^{1}$,
Lina Necib$^{7,8}$,
\newauthor
Claude-Andr\'e Faucher-Gigu\`ere$^{9}$,
Arpit Arora$^{3}$,
Pratik J. Gandhi$^{6}$
\\
$^{1}$Center for Computational Astrophysics, Flatiron Institute, 162 5th Ave., New York, NY 10010, USA\\
$^{2}$Department of Astronomy, Columbia University, 550 West 120th Street, New York, NY 10027, USA\\
$^{3}$Department of Physics $\&$ Astronomy, University of Pennsylvania, 209 S 33rd St., Philadelphia, PA 19104, USA\\
$^{4}$Institute for Computational Cosmology, Department of Physics, Durham University, South Road, Durham DH1 3LE, UK\\
$^{5}$Centre for Extragalactic Astronomy, Department of Physics, Durham University, South Road, Durham DH1 3LE, UK\\
$^{6}$Department of Physics and Astronomy, University of California, Davis, CA 95616, USA\\
$^{7}$The NSF AI Institute for Artificial Intelligence and Fundamental Interactions\\
$^{8}$Department of Physics and Kavli Institute for Astrophysics and Space Research, Massachusetts Institute of Technology,\\ 77 Massachusetts Ave, Cambridge MA 02139, USA\\
$^{9}$Department of Physics $\&$ Astronomy and CIERA, Northwestern University, 1800 Sherman Ave, Evanston, IL 60201, USA\\
$^{\dagger}$ Hubble Fellow.
}

\date{Accepted XXX. Received YYY; in original form ZZZ}

\pubyear{2015}

\usepackage[normalem]{ulem}

\begin{document}
\label{firstpage}
\pagerange{\pageref{firstpage}--\pageref{lastpage}}
\maketitle

\begin{abstract}
Observational studies are finding stars believed to be relics of the earliest stages of hierarchical mass assembly of the Milky Way (i.e., proto-Galaxy). In this work, we contextualize these findings by studying the masses, ages, spatial distributions, morphology, kinematics, and chemical compositions of proto-galaxy populations from the 13 Milky Way (MW)-mass galaxies from the FIRE-2 cosmological zoom-in simulations. Our findings indicate that proto-Milky Way populations: \textit{i}) can have a stellar mass range between $1\times10^{8}<\mathrm{M}_{\star}<2\times10^{10}~[\mathrm{M}_{\odot}]$, a virial mass range between $3\times10^{10}<\mathrm{M}_{\star}<6\times10^{11}~[\mathrm{M}_{\odot}]$, and be as young as $8 \lesssim \mathrm{Age} \lesssim 12.8$ [Gyr] ($1\lesssim z \lesssim 6$); \textit{ii}) are predominantly centrally concentrated, with $\sim50\%$ of the stars contained within $5-10$ kpc; \textit{iii}) on average show weak but systematic net rotation in the plane of the host's disc at $z=0$ (i.e., $0.25\lesssim\langle\kappa/\kappa_{\mathrm{disc}}\rangle\lesssim0.8$); \textit{iv}) present [$\alpha$/Fe]-[Fe/H] compositions that overlap with the metal-poor tail of the host's old disc; \textit{v}) tend to assemble slightly earlier in Local Group-like environments than in systems in isolation. Interestingly, we find that $\sim60\%$ of the proto-Milky Way galaxies are comprised by 1 dominant system ($1/5\lesssim$M$_{\star}$/M$_{\star,\mathrm{proto-Milky Way}}$$\lesssim4/5$) and $4-5$ lower mass systems (M$_{\star}$/M$_{\star,\mathrm{proto-Milky Way}}$$\lesssim1/10$); the other $\sim40\%$ are comprised by 2 dominant systems and $3-4$ lower mass systems. These massive/dominant proto-Milky Way fragments can be distinguished from the lower mass ones in chemical-kinematic samples, but appear (qualitatively) indistinguishable from one another. Our results could help observational studies disentangle if the Milky Way formed from one or two dominant systems.
\end{abstract}




\section{Introduction}

In the current accepted model for cosmology ($\Lambda$CDM), haloes grow by accumulating lower mass building blocks through a process commonly referred to as hierarchical mass assembly (\citealp[e.g.,][]{White1978,White1991}). During this process, the baryonic components (i.e., gas and stars that constitute observable galaxies) of haloes also undergo this mechanism of mass accumulation. This procedure is ubiquitous across the Universe and affects all galaxies. Thus, the accretion history of a galaxy is a pivotal dictating factor for its evolution and assembly of mass over time, and the clues for disentangling such intricate process are all contained in the stellar halo.

Our current picture for the formation of stellar haloes suggests that they form via a dual process. On the one hand, gas accretion from cosmic filaments drives secular evolution and in situ star formation. On the other, the accretion of different mass building blocks\footnote{A definition of building block and main branch systems is provided in Table~\ref{tab0}.}, each of which donate their gas and stars to the resulting larger mass host, contribute in mass to a given galaxy after becoming consumed. For the case of the Milky Way, early observational evidence has suggested that the former process dominates within the inner regions (i.e., $r\lesssim20-30$ kpc), whereas the latter dominates the outer regime (\citealp[e.g.,][]{Chiba2000,Carollo2007,Deason2011}). However, more recent observational results suggest that the in situ component may be a heated primordial disc (\citealp[e.g.,][]{Bonaca2017,Matteo2019,Belokurov2020}). This dual formation channel has also been somewhat shown to be the case in Andromeda (M31; \citealp[e.g.,][]{Ferguson2002,Brown2006, Escala2019, Escala2020}). The advent of detailed semi-analytic cosmological models and more recent detailed cosmological simulations have supported this hypothesis on a theoretical basis (\citealp{Bullock2005,Abadi2006,Bell2008,Font2008,Johnston2008,Cooper2010,Font2011,McCarthy2012,Amorisco2017, Khoperskov2022_a, Khoperskov2022_2}). However, whilst this picture may appear clear at current time, detailed spectroscopic, photometric, and astrometric observations of halo stars with \textit{Gaia} \citep[][]{gaia2022} and large stellar surveys are starting to poke holes in this framework.

Recent observational results have shown that the inner $\sim30$ kpc of the Milky Way's stellar halo is awash with debris from engulfed satellite systems. More specifically, it has been shown that in the Milky Way: $i$) the local stellar halo is dominated by the debris of a massive and (likely) ancient merged galaxy (\textit{Gaia}-Sausage/Enceladus; \citealp{Belokurov2018,Helmi2018,Haywood2018,Mackereth2019}), although see a counter-argument by \citet{Donlon2022}; $ii$) the innermost regions of the Galaxy ($r<5$ kpc) likely host the debris from an ancient and massive building block (Heracles; \citealp{Horta2021})\footnote{It has been postulated that globular clusters in the inner Galaxy could be linked to this system (\citealp[][]{Kruijssen2020,Forbes2020}).}; this system could be related to another population recently identified, referred to as the "proto-Milky Way" or Aurora (\citealp{Belokurov2022, Conroy2021, Rix2022}); $iii$) there are numerous smaller-mass halo substructures/streams postulated to be the debris from smaller mass accreted systems (see \citealp{Helmi2020} for a review). Along those lines, further out in the stellar halo, there are also streams from currently/recently disrupted satellites (e.g., Sagittarius dSph, \citealp{Ibata1994}; Cetus, \citealp{Newberg2009}; Orphan-Chenab, \citealp{Gillmair2006,Belokurov2007}) \footnote{And of course the Magellanic Clouds.}. All these different systems suggest that many components contribute to the formation of the Milky Way's stellar halo at all radii, in agreement with recent findings using cosmological simulations (\citealp[e.g.,][]{Khoperskov2022_2,Horta2023,Orkney2023}). 

The recent plethora of observational findings hint towards the epoch of formation for the Milky Way being early ($\gtrsim8$ Gyr). This implies that the major building blocks of the Galaxy are many gigayears old. While lower-mass/unmixed accretions are interesting for unveiling the recent mass assembly history of the Galaxy, or are useful in terms of potential measuring and dynamical studies, it is the early building blocks and accretion events that: $i$) constitute the bulk of the inner regions of the present day stellar halo (i.e., $r\lesssim30$ kpc); $ii$) supply the gas and baryonic material to fuel the formation of the disc; $iii$) dictate key episodes in the formation of the Milky Way. For this reason, observational studies have set out to investigate the properties of the oldest stellar populations in the Milky Way, and have reported chemical-kinematic evidence for the presence of stellar populations in the innermost regions of the Galaxy ---where one would expect the oldest stars formed in situ to inhabit (\citealp{Badry2018,Fragkoudi2020})--- that are distinguishable from the dominant bar/disc, and are likely to constitute the entirety, or part of, the ``proto-Milky Way''. This stellar population has been postulated to arise from the main progenitor system of the Milky Way, a major building block, or both simultaneously. Whilst the nature of this population is yet to be fully established, these findings have instigated the search for answers to additional pivotal questions when it comes to understanding the formation of the Galaxy: 
\begin{itemize}
\item \textit{When did the proto-Milky Way form?}
\item \textit{What constitutes the proto-Milky Way?}
\item \textit{Is it possible to distinguish the different systems that formed the proto-Milky Way from one another?}
\end{itemize}

\citet{Santistevan2020} set out to investigate the formation times and building blocks of Milky Way-mass galaxies in the FIRE-2 simulations, which is directly related to the formation of proto-Milky Way populations. In their work, the authors traced the origin of star particles formed at all times in different regions of the Milky Way-mass galaxies, and found that: $i$) Milky Way-mass galaxies typically assemble at $z \sim3-4$ (i.e., $11.6-12.2$ Gyr ago); $ii$) Milky Way-mass galaxies at $z=0$ are formed typically from $\sim100$ building blocks with M$_{\star} \gtrsim 10^{5}$ M$_{\odot}$; Milky Way-mass galaxies in Local Group environments typically assemble earlier than those in isolation, highlighting the importance of environment.

In this article, we set out to answer when and how proto-Milky Way systems may have assembled, and what are the (present day) observable properties of stellar populations that comprise a proto-Milky Way. To do so, we employ the thirteen Milky Way-mass galaxies from the \textit{Latte} and \textit{ELVIS} suites \citep{Wetzel2016} of high-resolution cosmological simulations from the Feedback In Realistic Environments (FIRE: \citealp{Hopkins2018}) project (Section~\ref{data}). We take these simulations and track the star particles belonging to all the resolvable luminous subhaloes, in order to identify the star particles belonging to all the galactic systems that coalesce to form a proto-Milky Way (Section~\ref{sec_proto_bbs}). Upon selecting these populations, in Section~\ref{sec_results} we go on to examine the present day observable properties of their stellar debris (namely, mass, age, spatial distribution, morphology, kinematics, and [$\alpha$/Fe] and [Fe/H] chemical compositions), in order to test if different systems constituting proto-Milky Way populations are distinguishable, and to provide clues on how to identify proto-Milky Way populations. We then discuss our results in the context of the current/future work in Section~\ref{sec_discussion}, provide the limitations and improvements to our work in Section~\ref{sec_limitations}, and list our concluding statements in Section~\ref{sec_conclusions}.

\section{Simulations}
\label{data}

We make use of thirteen Milky Way-mass galaxies from the \textit{Latte} \citep{Wetzel2016} and \textit{ELVIS} \citep{Garrison2019_mass} suites of FIRE-2\footnote{FIRE project website: \url{http://fire.northwestern.edu}} cosmological zoom-in simulations. In detail, \textit{Latte} is a suite of seven isolated Milky Way-mass galaxies, and \textit{ELVIS} is a suite of three Local Group-like pairs of Milky Way-mass galaxies. Both these suites of simulations were run with the FIRE-2 physics model \citep{Hopkins2018}, utilising the Lagrangian meshless finite-mass $N$-body gravitational plus hydrodynamics code \texttt{GIZMO}\footnote{\url{http://www.tapir.caltech.edu/~phopkins/Site/GIZMO.html}.} \citep{Hopkins2015}. FIRE-2 simulations model many radiative cooling and heating processes for gas, including free-free emission, photoionization/recombination, Compton scattering, photoelectric, metal-line, molecular, fine-structure, dust-collisional, and cosmic-ray heating across a temperature range of 10 - 10$^{10}$K. These simulations also include the spatially uniform, redshift-dependent, cosmic UV background from \citet{Faucher2009}, for which HI reionization occurs at $z_{\mathrm{reion}}$ $\sim$ 10. Moreover, FIRE-2 self-consistently generates and tracks 11 chemical abundance species (namely,  H, He, C, N, O, Ne, Mg, Si, S, Ca, and Fe), including sub-grid diffusion of these abundances in gas via turbulence (\citealp{Hopkins2016,Su2017,Escala2018}), as well as enrichment from core-collapse supernovae, supernovae type Ia, and stellar winds.

The \textit{Latte} suite has an initial baryonic mass resolution of 7,100 M$_{\odot}$ for gas particles, whereas the \textit{ELVIS} suite has an initial baryonic mass resolution of $3,500-4,000$ M$_{\odot}$. \textit{Latte} uses a fixed resolution of 35,000 M$_{\odot}$ for dark matter particles, whereas \textit{ELVIS} uses a fixed resolution of 19,000 M$_{\odot}$. In both \textit{Latte}/\textit{ELVIS} star formation occurs in gas that is self-gravitating, Jeans-unstable, cold (T $<$ 10$^{4}$K), dense ($n$ $>$ 1,000 cm$^{-3}$), and molecular (following \citealt{Krumholz2011}). Each star particle inherits the mass and chemical abundance composition of its progenitor gas particle, and represents a single stellar population with a \citet{Kroupa2001} initial mass function. This population evolves according to the \texttt{STARBURST} v7.0 models \citep{Leitherer1999}, so the star particle decreases in mass with time as the most massive stars die. By the present day most star particles have a mass of $4,000-5,000$ M$_{\odot}$. Stellar evolution gives rise to localized feedback at the location of each star particle, including core-collapse and Ia supernovae, mass loss from stellar winds, and radiation, including radiation pressure, photoionization, and photo-electric heating. Implementation of these processes follows the descriptions in \citet{Hopkins2018} and \citet{2018MNRAS.477.1578H}. 

The \textit{Latte}/\textit{ELVIS} suites were generated within periodic cosmological boxes of lengths 70.4--172 Mpc using the code \texttt{MUSIC} \citep{hahn2011}, and constructing cosmological zoom-in initial conditions for each simulation at $z$ $\simeq$ 99. Each simulation has 600 snapshots saved down to $z$ = 0, spaced every $\simeq$ 25 Myr. All simulations assume flat $\Lambda$-CDM cosmology with parameters consistent with \citet{Plank2020}. More specifically, the \textit{Latte} suite (excluding m12r and m12w) used $\Omega_{\mathrm{m}}$ = 0.272, $\Omega_{\mathrm{b}}$ = 0.0455,
$\sigma_{\mathrm{8}}$ = 0.807, $n_{\mathrm{s}}$ = 0.961, $h$ = 0.702. The m12r and m12w halos were selected specifically because they host an LMC-mass satellite galaxy, and they adopted more up-to-date initial conditions from \citet{Plank2020} compared to the rest of the \textit{Latte} suite: $h$ = 0.68, $\Omega_{\mathrm{\Lambda}}$ = 0.31, $\Omega_{\mathrm{m}}$ = 0.31, $\Omega_{\mathrm{b}}$ = 0.048,
$\sigma_{\mathrm{8}}$ = 0.82, $n_{\mathrm{s}}$ = 0.961. For the \textit{ELVIS} suite, Thelma $\&$ Louise and Romulus $\&$ Remus both used the same cosmology as in the original \textit{ELVIS} dark matter only suite: $\Omega_{\mathrm{m}}$ = 0.266, $\Omega_{\mathrm{b}}$ = 0.0449, $\sigma_{\mathrm{8}}$ = 0.801, $n_{\mathrm{s}}$ = 0.963, $h$ = 0.71. Conversely, Romeo $\&$ Juliet used the same cosmology as m12r/m12w. The post-processing is done using \texttt{gizmo analysis} \citep[][]{Wetzel2020a} and \texttt{halo analysis} \citep[][]{Wetzel2020b}. Furthermore, dark matter particles in each snapshot are processed with \texttt{Rockstar} \citep[][]{Behroozi2013} to produce halo catalogs. For further details of how the \textit{Latte} and \textit{ELVIS} suites were generated, we refer the reader to \citet{Wetzel2016} and \citet{Garrison2019_mass} and references therein, respectively.

The resolution of this suite of simulations enables luminous subhalos to be well resolved even near each Milky Way-like galaxy. It also resolves the formation of tidal streams from satellite galaxies down to approximately $10^{8}$M$_{\odot}$ in total mass or $10^{6}$M$_{\odot}$ in stellar mass (at $z=0$), similar to that of the Milky Way's ``classical" dwarf spheroidals (\citealp[e.g.,][]{Panithanpaisal2021,Cunningham2022,Shipp2022,Horta2023}).

The  properties of the host galaxies in \textit{Latte} show broad agreement with the Milky Way, including the stellar-to-halo mass relation \citep[][]{Hopkins2018}, stellar halos (\citealp[][]{Bonaca2017, Sanderson2018,Sanderson2020}), and the radial and vertical structure of their disks (\citealp[][]{Ma2017,Bellardini2021}). Moreover, the satellite populations of these simulation suites have also been demonstrated to agree with several observed properties, such as: the mass and velocity dispersions (\citealp[][]{Wetzel2016,Garrison2019_dispersions}); star formation histories \citep[][]{Garrison2019_mass}; and radial distributions \citep[][]{Samuel2020}. Despite great similarities in the properties of the satellite galaxies in \textit{Latte} and that of the observed satellites around the Milky Way, it has been shown that the former are generally too metal-poor when compared to the latter (\citealp[][]{Escala2018,Wheeler2019,Panithanpaisal2021}). This could be in part because of the assumed delay time distributions assumed for SNIa \citep{Gandhi2022}, or because of the lack of modeling of Pop III stars. For this work, we emphasize that we do not require quantitative agreement with simulated and observed abundances. The relations found in this paper between (luminous) haloes and their respective chemical abundances should be treated qualitatively, and are intended for use within the simulations only. However, although the \emph{normalization} of the various abundances is not always in good agreement with observations, we expect the \emph{trends} we identify in this paper to be robust.

\section{Defining a proto-Milky Way}
\label{sec_proto_bbs}

\begin{table*}
	\centering
	\caption{List of definitions used in this article. See also Fig~\ref{formdist}.}
	\begin{tabular}{lccr} 
		\hline
		    Name & Definition\\
		\hline
            $t_{\mathrm{MR_{3:1}}}$ & Lookback time in the simulation we define the proto-Milky Way to emerge, defined as the time \\& the main branch reaches a stellar mass ratio of 3:1 with the second most massive luminous subhalo \\& in the simulation volume (i.e., $\sim2$ Mpc or up to $\sim6~R_{\mathrm{vir}}$ for \textit{Latte}, and approximately double for \textit{ELVIS}).\\
            Main branch &  Star particles formed in the main halo (as defined by the simulation) before $t_{\mathrm{MR_{3:1}}}$ \\
            Building block &  Star particles formed in luminous halos that join with the main branch halo before $t_{\mathrm{MR_{3:1}}}$ \\
            Proto-Milky Way &  Star particles formed in the main branch and building block systems, that coalesce into a single galaxy before $t_{\mathrm{MR_{3:1}}}$ \\
            \hline
        \end{tabular}
	\label{tab0}
\end{table*}

\begin{figure*}
\centering
\includegraphics[width=\textwidth]{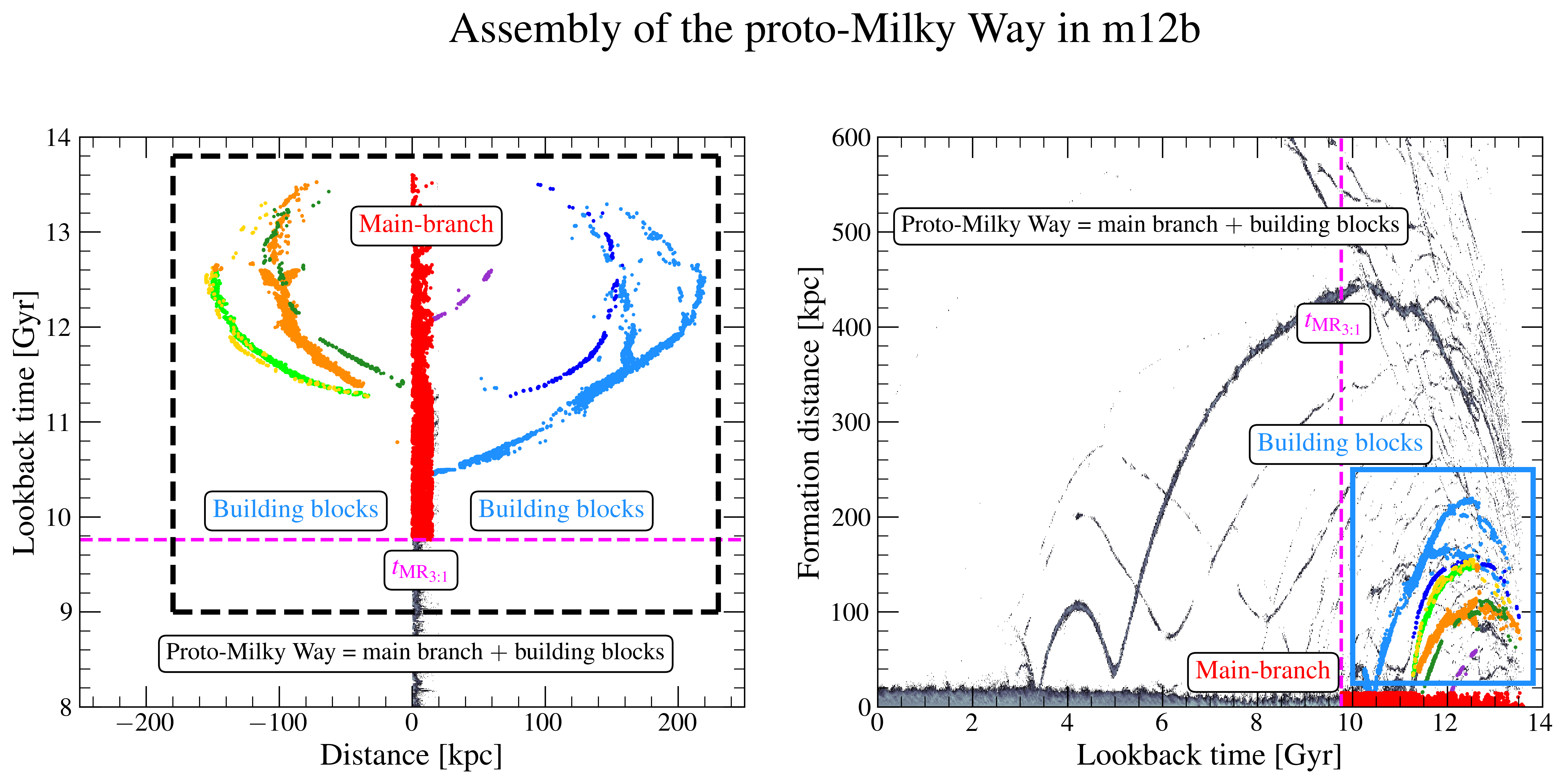}
\caption{\textit{Left}: Diagram of the merger tree of the m12b simulation in \textit{Latte} up to a lookback time of 8 Gyr. In this work, a proto-Milky Way is defined as the amalgamation of the main branch halo (red) in a simulation \textit{plus} all the building blocks (other colours) that coalesce onto it before $t_{\mathrm{MR_{3:1}}}$ (dashed magenta line, see Table~\ref{tab0} for details). \textit{Right}: Distance at which a star particle in the simulation is formed w.r.t. the centre of the main host as a function of lookback time (i.e., age) for the m12b simulation in \textit{Latte}. Highlighted in red are the star particles associated with the main branch, and as other colours the star particles associated with the resolvable building blocks (i.e., luminous subhaloes) that join with the main branch before $t_{\mathrm{MR_{3:1}}}$. By tracking each system over time, we are able to identify the star particles associated with all the proto-Milky Way fragments.}
\label{formdist}
\end{figure*}

Before performing any analyses, it is imperative that we make clear our definition of a proto-Milky Way. To do so, we set out to answer the following two questions:\\

\textit{1) When does a proto-Milky Way form?}\\

\textit{2) What constitutes a proto-Milky Way?}\\

For this work, we define the time in which a system becomes a proto-Milky Way when the main halo in the zoom-in region of the simulation box ($\sim2$ Mpc side box or up to $\sim$6 $R_{\mathrm{vir}}$ for \textit{Latte}, and approximately double for \textit{ELVIS}) reaches a stellar mass ratio of 3:1 with the second most luminous halo\footnote{This second most luminous halo is by definition not in our sample of systems that comprise the proto-Milky Way.} ($t_{\mathrm{MR_{3:1}}}$). We note that for the \textit{ELVIS} galaxies, as there are two main haloes in each simulation, we take the ratio of stellar mass between the two main host haloes and the third most luminous subhalo within the simulation. Albeit arbitrary, we argue that this definition is intuitively sensible as: $i$) it defines a point in time in which a system is dominant in its environment; $ii$) is quantifiable and repeatable across different simulations. It has also been reasoned to be a sensible way of pinpointing the time in which a galaxy becomes dominant \citep{Santistevan2020}. \footnote{We note that \citet{Santistevan2020} also use another sensible choice, defined as the time in which the main halo begins forming the majority of its present day stars, signifying a dominant star formation episode of sustainable growth from an \textit{in situ} channel.} Given our definition, it is then straight forward to answer question \textit{2)}; the proto-Milky Way is defined as the stellar population resulting from the amalgamation between the main branch (i.e., the halo tracing the formation of the main host in the simulation) and building blocks (i.e., all the haloes that join with the main branch) before $t_{\mathrm{MR_{3:1}}}$. Fig \ref{formdist} shows a schematic of our definitions, where the main branch is shown in red and building blocks as other colours. Moreover, a list of definitions is provided in Table\ref{tab0}; Table~\ref{tab1} displays the  $t_{\mathrm{MR_{3:1}}}$ times, the number of systems that coalesce to form a proto-Milky Way (given our definitions), and the stellar/virial masses for each proto-Milky Way system.

\begin{table*}
	\centering
	\caption{List of approximate times in the simulation in which we define the proto-Milky Way to form ($t_{\mathrm{MR_{3:1}}}$), as well as the number of events (i.e., main branch+building blocks) that constitute the proto-Milky Way ($n_{\mathrm{systems}}$), and the stellar and virial mass of the proto-Milky Way at $t_{\mathrm{MR_{3:1}}}$. We note here that the number of building blocks we find is bounded by our choice to track only systems with 150 star particles or more (see \citet{Santistevan2020} for a comparison when tracking lower mass systems). We also list the corresponding references for each of the \textit{Latte}/\textit{ELVIS} simulated haloes: A. \citet{Hopkins2018}; B. \citet{Garrison2019_dispersions}; C. \citet{Garrison2019_mass};, D. \citet{Garrison2017}; E. \citet{Wetzel2016}; F. \citet{Samuel2020}. These (lookback) times should be interpreted as ages, where 13.8 Gyr is old and 0 Gyr is young. The average values for $n_{\mathrm{systems}}$ have been rounded to the closest integer. The $t_{\mathrm{MR_{3:1}}}$ agree well with the results found in \citet{Santistevan2020} (see their Table 1).}
	\begin{tabular}{lcccccccr} 
		\hline
		Host & $t_{\mathrm{MR_{3:1}}}$ [Gyr] & $t_{\mathrm{MR_{3:1}}}$ [$z$] & $n_{\mathrm{systems}}$ & M$_{\star_{3:1}}$ [$\times$10$^{9}$ M$_{\odot}$] & M$_{\mathrm{vir}_{3:1}}$ [$\times$10$^{11}$ M$_{\odot}$]\\
		\hline
            (\textit{isolated})\\
            m12b$^{\mathrm{C}}$ & 9.76& 1.59& 8 & 8.80& 4.43\\
            m12c$^{\mathrm{C}}$ & 9.05& 1.32&5 & 3.56& 4.93\\
            m12f$^{\mathrm{D}}$ & 12.15& 3.77&4 & 0.38& 2.86\\
            m12i$^{\mathrm{E}}$ & 11.80& 3.18&6& 0.35& 1.12\\
            m12m$^{\mathrm{A}}$ & 9.94& 1.68&5& 3.69& 5.00\\
            m12r$^{\mathrm{F}}$ & 8.05& 1.02&7& 4.29& 2.61\\  
            m12w$^{\mathrm{F}}$ & 11.60& 2.92 &9 & 0.35& 1.19\\
            \hline
            (\textit{pairs})\\
            Juliet$^{\mathrm{C}}$ & 12.55& 4.72& 4 & 0.31&0.93\\
            Louise$^{\mathrm{C}}$ & 11.90& 3.32& 4 & 0.99&1.02\\
            Remus$^{\mathrm{B}}$ & 7.90& 0.98& 9 & 20.14& 6.41\\
            Romeo$^{\mathrm{C}}$ & 12.81& 5.70& 4 & 0.12&0.33\\
            Romulus$^{\mathrm{B}}$ & 10.00& 1.70& 6 & 6.85& 3.40\\
            Thelma$^{\mathrm{C}}$ & 12.40& 4.29& 2 & 0.11& 0.78\\
            \hline
            Average (isolated) & 10.33& 1.87 & 6 & 3.06&3.16\\
            Average (pairs)  &11.26& 2.56& 5& 4.75&2.14\\
            Average (total) &10.76 &2.15 & 6&3.90&2.69\\
            \hline
    \end{tabular}
	\label{tab1}
\end{table*}

Furthermore, in order to track halos with a high enough number of star particles to be resolvable in \textit{Latte}/\textit{ELVIS}, we choose to only track systems with 150 star particles or more, using a similar technique to previous studies (e.g., \citealp{Necib2019,Panithanpaisal2021,Horta2023}). More specifically, we track the evolution of each system (i.e., the star particles) at every snapshot in the simulation over time until $t_{\mathrm{MR}_{3:1}}$ or until the system no longer exists (i.e., it has merged with the main branch system) using the \texttt{halo} catalogues, instead of resorting to the merger trees. This allows us to identify all the star particles in every halo, and to assign star particles to individual building blocks. Our choice to only track systems with 150 star particles or more leads to a minimum stellar mass of M$_{\star}\sim1\times10^{6}$M$_{\odot}$ for subhaloes in \textit{Latte}, and M$_{\star}\sim5\times10^{5}$M$_{\odot}$ for \textit{ELVIS}, given the slight differences in particle resolution between these two sets of simulations. For more information on the limitations of this choice, see Section~\ref{sec_limitations}. 

\subsection{When does a proto-Milky Way form?}
Given our assumptions, we find that on average, proto-Milky Way populations are old (see Table~\ref{tab1}). These values range from as late as $t_{\mathrm{MR_{3:1}}} = 8.05$ Gyr ($z\sim1$) to as early as $t_{\mathrm{MR_{3:1}}} = 12.8$ Gyr ($z\sim6$). We find that overall the proto-Milky Way systems can be grouped into three main camps: an early forming group, an intermediate forming group, a late forming one. The difference in formation times of these Milky Way-mass galaxies has also been divided into three similar groups based on the time in which their stellar discs settle \citep{McCluskey2023}, and based on the transition from spheroids to thick and thin discs \citep{Yu2023}. These results are also consistent with the results from \citet{Santistevan2020} when they use the same definition as used here ($t_{\mathrm{MR_{3:1}}}$) (though we note that they find that Milky Way-mass galaxies emerge around $z\sim3-4$ when using a definition based on in-situ star formation).

\subsection{What constitutes a proto-Milky Way?}

\begin{figure}
\centering
\includegraphics[width=\columnwidth]{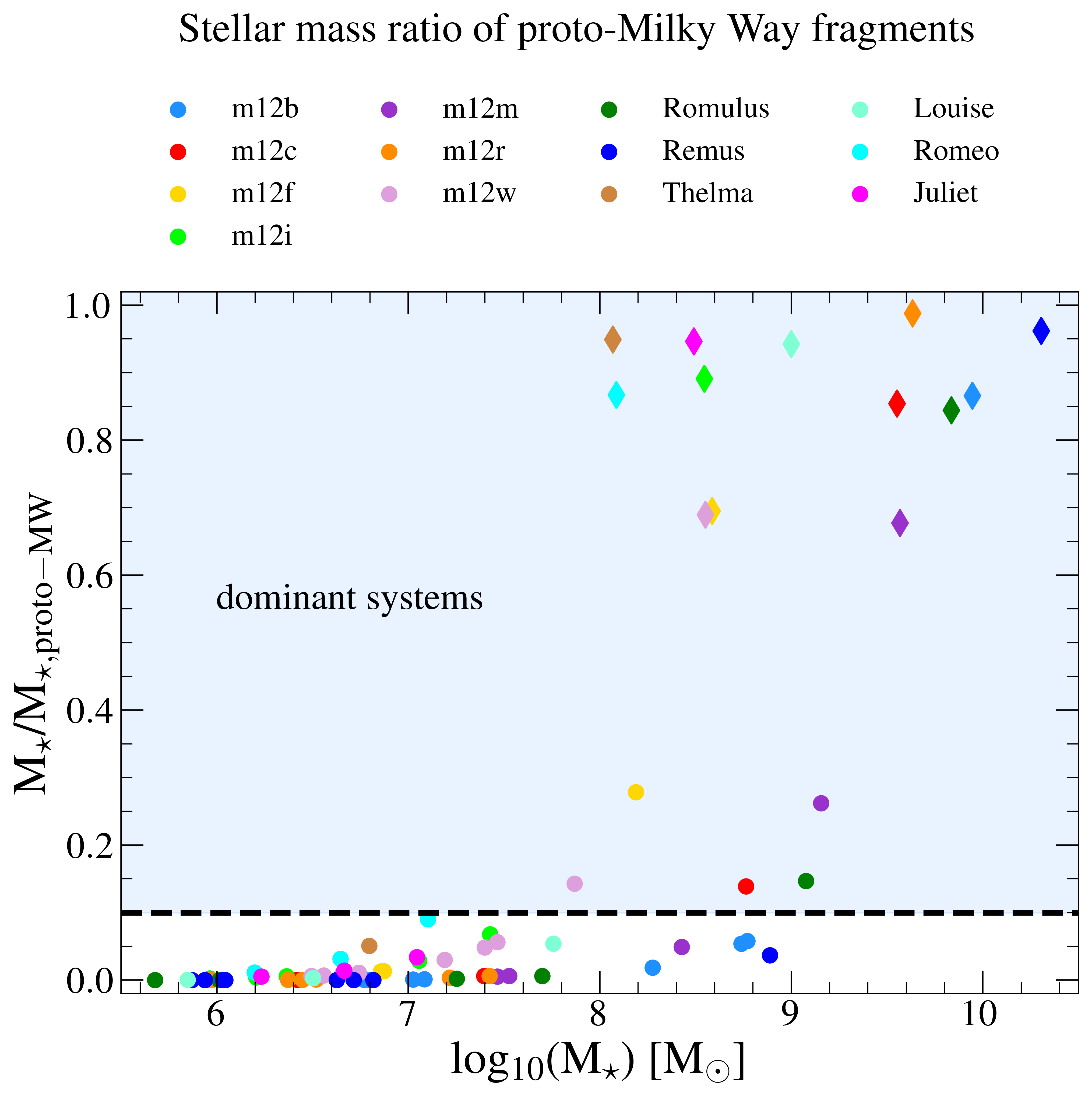}
\caption{Stellar mass ratio between each main branch (diamonds) or building block (circles) and the proto-Milky Way (i.e., main branch$+$building blocks) at $t_{\mathrm{MR_{3:1}}}$ as a function of the stellar mass of each main branch/building block. There are five clear cases (namely, m12c, m12f, m12m, m12w, and Romulus) which have a building block on the order of $\gtrsim$1:5 mass ratio with the proto-Milky Way population (i.e., blue shaded region). This indicates that $\sim$40$\%$ of the thirteen proto-Milky Way systems studied have a massive/dominant building block in addition to the dominant main branch halo.}
\label{fig_summary_mass}
\end{figure}

Fig~\ref{fig_summary_mass} shows the stellar mass ratio between every main branch (diamonds) and building block (circles) with the overall proto-Milky Way population as a function of the stellar mass of each main branch/building block at $t_{\mathrm{MR}_{3:1}}$.  In this diagram, there are proto-Milky Way populations formed by one clearly dominant halo (e.g., m12i), and populations formed by two main systems (namely, the main branch and a massive building block with a mass ratio of greater than 1:5; e.g., m12m). 
To guide the eye, in Fig~\ref{fig_summary_mass} we have highlighted the dominant systems as the shaded region, above the dashed line. Here, there are five building block systems (circles) with mass ratios of $\gtrsim1:5$ in addition to the dominant main branch (diamonds). Thus, for $\sim40\%$ of our sample (5/13 galaxies), the proto-Milky Way system is comprised of two dominant populations. Conversely, the other $\sim60\%$ (8/13 galaxies) host only one dominant main branch system. 

\begin{figure}
\centering
\includegraphics[width=\columnwidth]{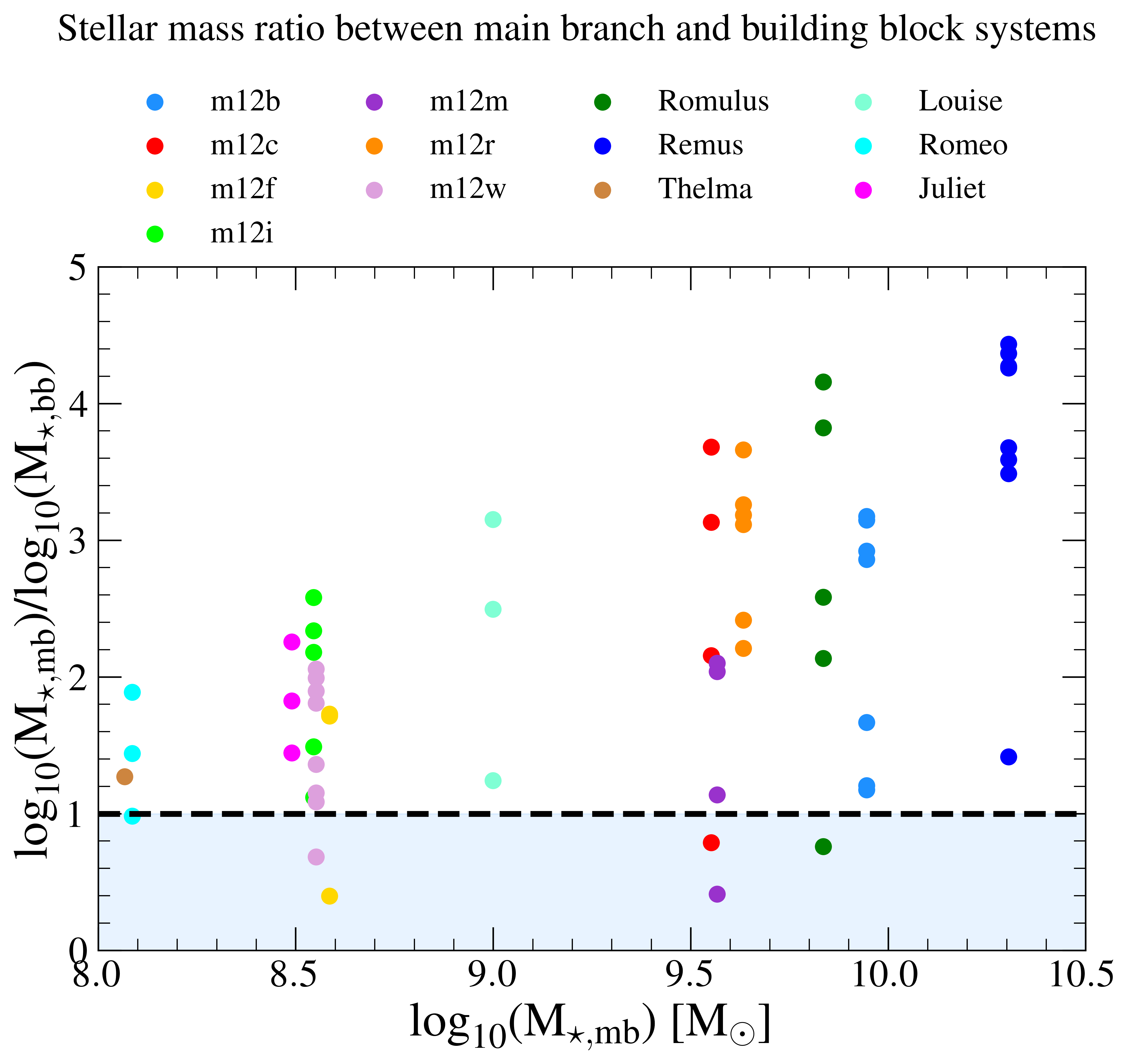}
\caption{Ratio of the stellar mass between main branch progenitors and their building block counterparts as a function of the main branch progenitor's stellar mass for each Milky Way-mass simulation at $t_{\mathrm{MR}_{3:1}}$. Five out of thirteen proto-Milky Way's host a building block of similar mass to the main branch progenitor (blue shaded region).}
\label{mass-ratios}
\end{figure}

Given this finding, it is interesting to ask \textit{what fraction of the proto-galaxy's (stellar) mass is contributed by the main branch/building block progenitors?} To answer this question, in Fig~\ref{mass-ratios} we show the mass difference between each main branch progenitor and its counterpart building blocks as a function of the mass of the main branch progenitor. We find that the majority of the building blocks are of lower mass when compared to the main branch progenitor, on the order of one to four orders of magnitude difference, with a stellar mass ranging between $5\times10^{5}<\mathrm{M}_{\star}<1\times10^{8}$ M$_{\odot}$. However, Fig~\ref{mass-ratios} confirms the findings from Fig~\ref{fig_summary_mass}, highlighting that for 5/13 galaxies in our sample, in addition to the main branch system, there is a building block whose stellar mass is of the same order of magnitude.

Our findings suggest that on average the proto-galaxy of a Milky Way-mass halo is comprised of five to six systems with M$_{\star}\gtrsim5\times10^{5}$M$_{\odot}$ and one to two systems with M$_{\star}\gtrsim1\times10^{8}$M$_{\odot}$\footnote{However, these results are subject to our choice to only study systems with 150 star particles. If one was to lower this mass limit and was to look at all systems that coalesce with the main branch before $z=0$ \citep[e.g.,][]{Santistevan2020}, this number grows significantly, up to $\sim$100 building blocks.}. In the following sections we set out to study the chemical-kinematic properties of these systems at $z=0$ to see if the different systems that form proto-Milky Way populations can be distinguished.

\section{Results}
\label{sec_results}

In this section we consider each main branch and building block population individually in order to examine how the mass, age, spatial distribution, morphology, kinematics, and abundance patterns of these systems differ. We also aim to examine the role of environment by comparing isolated Milky Way-mass haloes with ones in Local Group environments. 

\subsection{Stellar mass and age}
\label{sec_massage}

\begin{figure*}
\centering
\includegraphics[width=\textwidth]{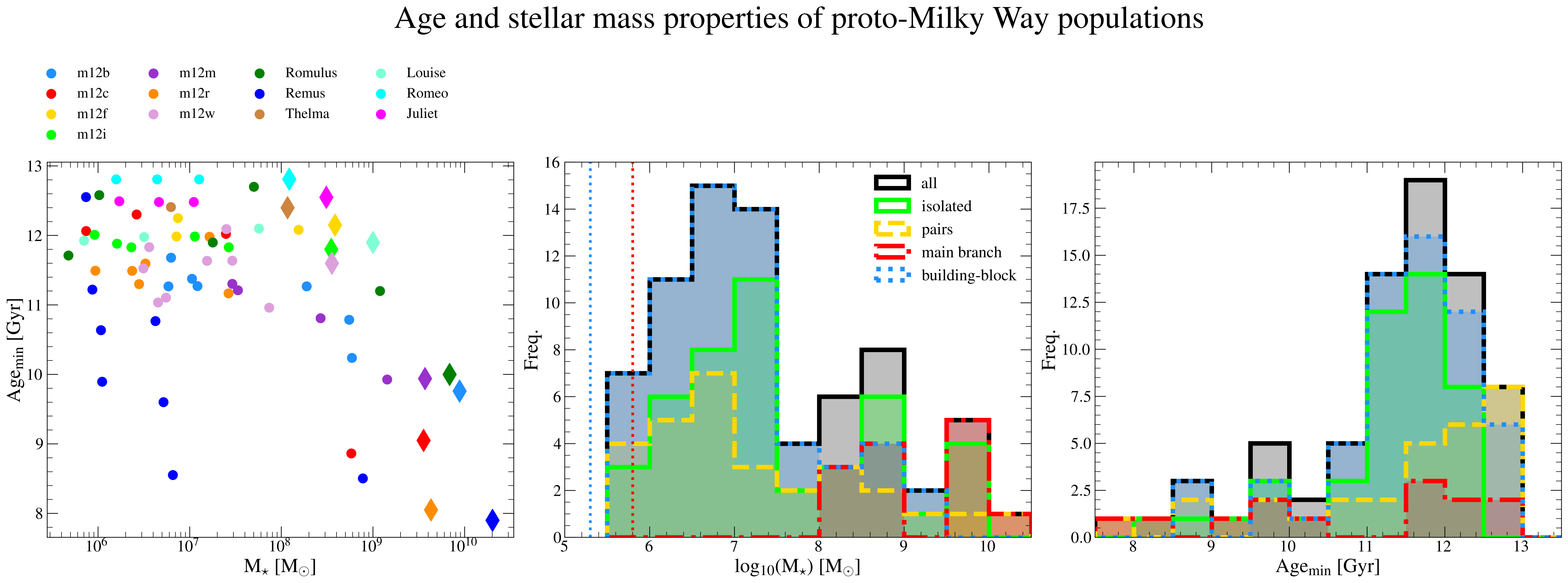}
\caption{\textit{Left}: minimum star particle age (i.e., the youngest star particle) in each main branch (diamond) and building block (circle) event as a function of its stellar mass for all thirteen Milky Way-mass haloes. \textit{Middle}: histogram of the stellar mass of each system, split into environment and main branch/building block populations. The vertical dotted lines demark the minimum possible stellar mass of a halo in the simulation given our choice to track luminous subhalos with 150 star particles or more and the simulations resolution, for \textit{Latte} (red) and and \textit{ELVIS} (blue). \textit{Right}: histogram of the minimum star particle age, illustrated as in the middle panel. In each panel, the label ``pairs'' corresponds to the systems studied in \textit{ELVIS}, and in a Local Group-like environment at $z=0$.}
\label{fig_agemass}
\end{figure*}

We begin by examining the stellar mass and the minimum age of a star particle in each main branch and building block separately, shown in Fig~\ref{fig_agemass}. We choose to study the minimum age instead of the mean age as we believe that it is a better constraint for the time in which a system quenches star formation, and thus ceases to evolve. The questions we aim to answer are: \textit{What age are the systems that form a proto-Milky Way?} \textit{How massive are the main branch and building block systems that constitute a proto-Milky Way?}

The left panel of Fig~\ref{fig_agemass} shows the relationship between the minimum star particle age and the stellar mass at $z=0$ of the main branch (diamonds) and building blocks (circles) from each simulation. The middle and right panels show histograms for stellar mass and minimum star particle age, respectively. Here, we show the frequency of all systems together in black, isolated haloes (i.e., \textit{Latte}) in green, systems in Local Group environments (i.e., \textit{ELVIS}) in yellow, main branch systems in red, and building blocks in blue. We recall that the resolution of the \textit{Latte} and \textit{ELVIS} suites are different (see Section~\ref{data} and the figure caption for details).

\begin{table*}
	\centering
	\caption{Summary of the mean, median, and 1-$\sigma$ values for the stellar mass and minimum star particle age for our sample of events comprising proto-Milky Way systems.}
	\label{tab:age-mass}
	\begin{tabular}{lcccccr} 
		\hline
		sample & $\langle$log$_{10}$(M$_{\star})$$\rangle$ [M$_{\odot}$] & \textit{med}(log$_{10}$(M$_{\star}$)) [M$_{\odot}$] & $\sigma_{\log_{10}(\mathrm{M}_{*})}$ [M$_{\odot}$] &$\langle$age$_{\mathrm{min}}$$\rangle$ [Gyr] & \textit{med}(age$_{\mathrm{min}}$) [Gyr] & $\sigma_{\mathrm{age_{min}}}$ [Gyr]\\
		\hline
		proto-Milky Way (all) &   7.39  &  7.08  & 1.17  &  11.36 & 11.64  & 1.16\\
            \hline
            \textit{Environmental difference}\\
		proto-galaxy (pairs) &  7.21 &  6.79  & 1.25  & 11.50  & 11.97  & 1.41\\
		proto-galaxy (isolated) &  7.50  &  7.31  & 1.09  &  11.26 &  11.51 & 0.94\\
            \hline
           \textit{ Classification difference}\\
            main branch & 9.09  &  8.99  & 0.71  & 10.76  & 11.60  & 1.66\\
		building block & 7.02 &  6.81  & 0.88  & 11.10  & 11.20  &1.31\\
            \hline
            \textit{All}\\
		main branch (pairs) & 8.96 & 8.74 & 0.85& 11.26& 12.15& 1.76\\
		main branch (isolated) & 9.19 & 9.55 & 0.56& 10.33& 9.94 & 1.43\\
		building block (pairs) & 6.75 & 6.64 & 0.88& 11.57& 11.97&1.29\\
		building block (isolated) & 7.178 & 7.05 & 0.84& 11.43& 11.52&0.68\\
            \hline
	\end{tabular}
\end{table*}

From inspection of the middle and right panels of Fig~\ref{fig_agemass}, we find that, overall (black solid line), the mean stellar mass of the subhaloes that comprise proto-Milky Way systems is $\langle$M$_{\star}$$\rangle$$\sim$4$\times$10$^{7}$M$_{\odot}$, and the mean minimum age is $\langle$age$_{\mathrm{min}}$$\rangle$$\sim$11.4 Gyr (or $\langle$$z$$\rangle$$\sim$2.8). However, the range in these parameters is significantly large, and for both cases, we find that the distribution is skewed (or for the case of the stellar mass, even possibly bimodal). Given our set of choices and assumptions, we find that the stellar masses for all these systems ranges from $\sim$5$\times$10$^{5}$$<$M$_{\star}$$<$2$\times$10$^{10}$M$_{\odot}$, and the minimum age can reach anything between $\sim$8$<$age$_{\mathrm{min}}$$<$13 Gyr (or $\sim$0.7$<$$z$$<$6.5). Furthermore, when splitting the distributions by environment and comparing the haloes that live in isolated environments from those that live in a Local Group-like environment, we find that the mean values of the stellar masses of systems in Local Group-like environments is slightly lower than for systems in isolation. The difference in the median stellar mass is on the order of half a magnitude. Furthermore, there are also some slight differences in the distribution of the minimum star particle age, whereby the pairs tend to favour older populations. Statistically, the mean values show no clear differences ($\langle$age$_{\mathrm{min, pairs}}$$\rangle$ = 11.5 Gyr and $\langle$age$_{\mathrm{min, isolated}}$$\rangle$ = 11.26 Gyr). However, the median differs by approximately 0.5 Gyr for both the stellar mass and minimum age (see Table~\ref{tab:age-mass}).

When comparing the main branch populations (red) with their building block counterparts (blue), we see substantial differences in their stellar masses. Specifically, we find that main branch systems tend to be more massive, approximately two orders of magnitude larger (i.e., $\langle$M$_{\star,\mathrm{mb}}$$\rangle = 1\times10^{9}$M$_{\odot}$ and $\langle$M$_{\star,\mathrm{bb}}$$\rangle = 1\times10^{7}$M$_{\odot}$). However, we note that some building blocks do reach high stellar masses, on the order of M$_{\star}\sim2\times10^{9}$M$_{\odot}$, making the range in the distribution for building blocks much larger than for main branch debris. We see similar differences in the mean and spread of the minimum star particle age between main branch and building blocks debris as we did when comparing environment, on the order of $\sim0.4$ Gyr. 

Our results thus imply that proto-Milky Way populations are predominantly comprised by one or two major systems of similar mass to the LMC (i.e., M$_{\star}\sim1\times10^{9}$M$_{\odot}$) and 3--5 smaller mass building blocks of approximately $\sim2$ orders of magnitude smaller in stellar mass (i.e., M$_{\star}\sim4\times10^{7}$M$_{\odot}$). The massive systems typically arise from the main branch progenitor, and in the case where there are two dominant systems, also from a massive building block.

The left panel of Fig~\ref{fig_agemass} shows the distribution of minimum star particle age (i.e., the youngest) as a function of stellar mass for each event in each Milky Way-mass halo separately, where here we distinguish main branch populations (diamonds) and building blocks (circles). There is a relation between the stellar mass of a system and the minimum star particle age, whereby a system's minimum stellar age decreases with increasing stellar mass. This is likely because those systems that are forming stars for longer are able to build up more mass (a similar argument as for the mass-metallicity relation of galaxies, see also Figure 3 of \citealp{Cunningham2022}). These results from this section are summarised in Table~\ref{tab:age-mass}.

\subsection{Spatial distribution and morphology}
\label{sec_spatial}

\begin{figure*}
\centering
\includegraphics[width=\textwidth]{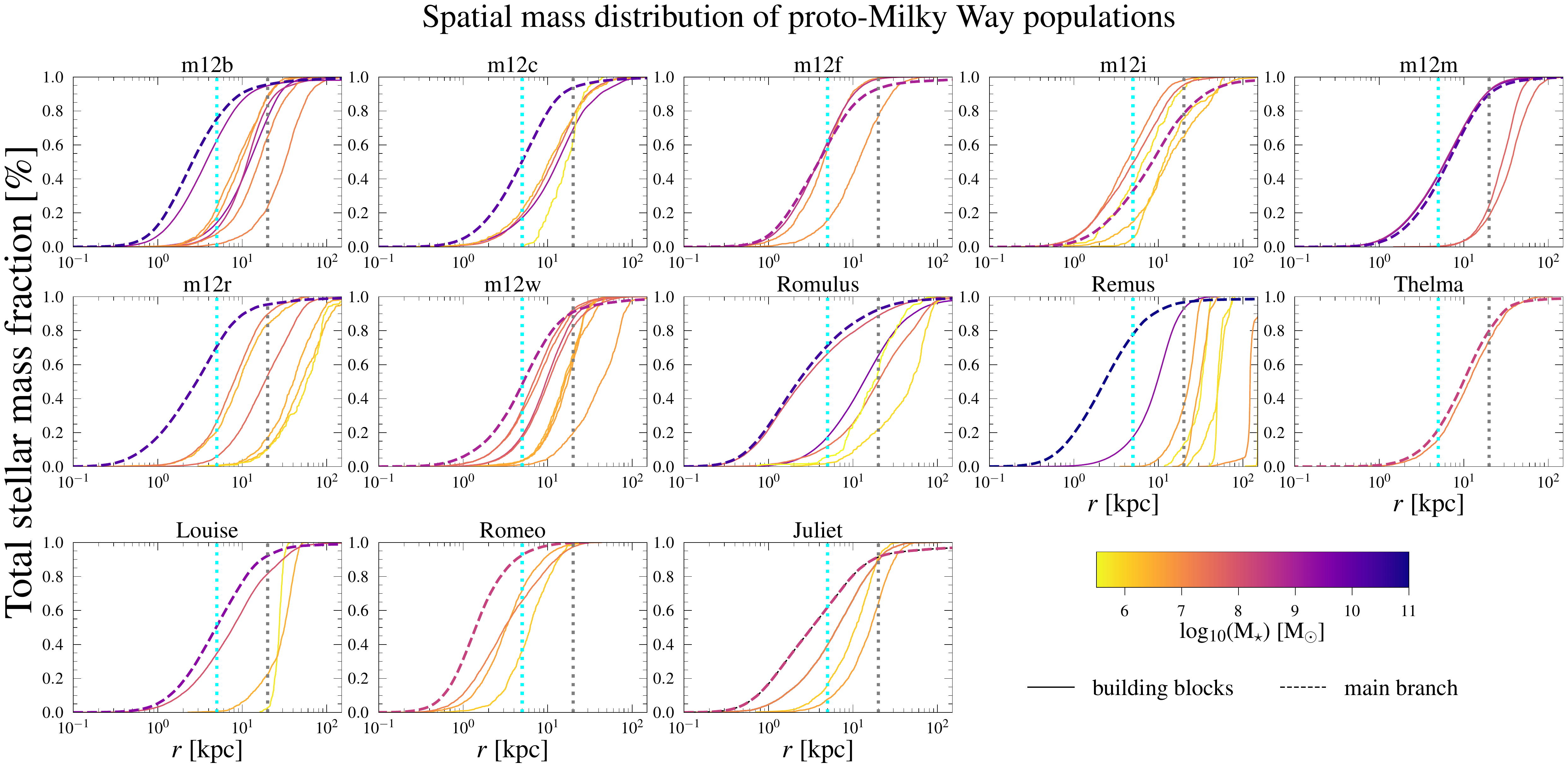}
\caption{Total stellar mass fraction as a function of spherical radius (at $z=0$) for all the main branch (dashed) and building block (solid) events in each Milky Way-mass halo. Each profile is colour coded by the systems' respective stellar mass. The vertical dashed lines mark 5 kpc (cyan) and 20 kpc (gray). The majority of the proto-Milky Way material (i.e., main branch and massive building blocks) are contained within a small spatial volume, close to the host's centre. However, there is a range of spatial profiles for lower mass building blocks.}
\label{radii}
\end{figure*}

We now set out to examine the spatial distribution of the different contributors to a proto-Milky Way. In doing so, we aim to tackle the following questions: \textit{Where is most of the mass contained in the main host at $z=0$?} \textit{What is the morphology of the debris at $z=0$?}

Fig~\ref{radii} characterizes where mass is deposited by showing the total (cumulative) mass fraction of the debris of every main branch (dashed) and building block (solid) event comprising the proto-Milky Way systems as a function of present day spherical radius from the centre of the Milky Way-mass host. Each main branch/building block event is colour coded by their respective stellar mass. To guide the readers eye, we also plot a vertical line at 5 kpc (cyan) and 20 kpc (grey).

\begin{figure}
\centering
\includegraphics[width=\columnwidth]{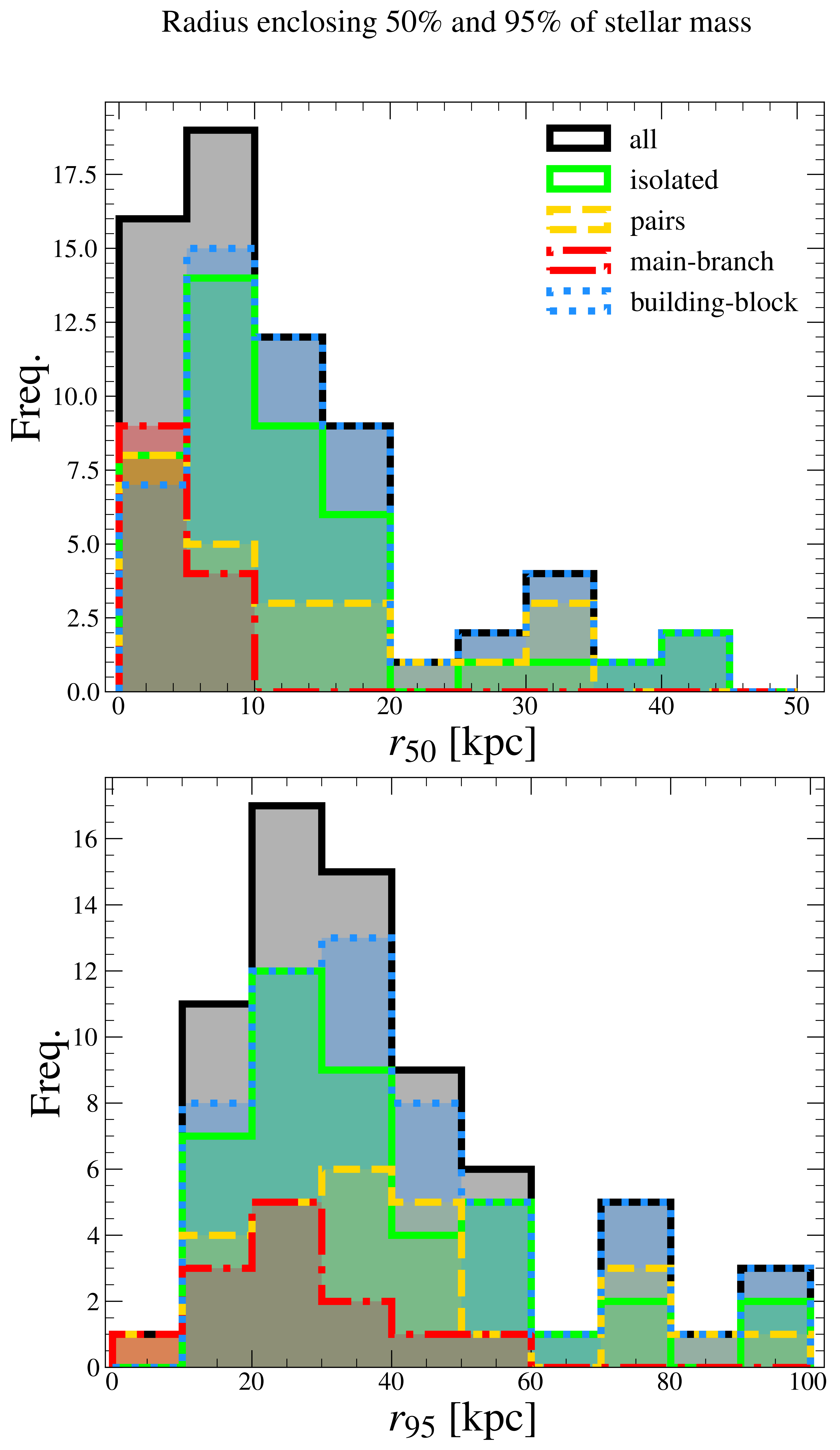}
\caption{Histograms of the 50$^{th}$ (top) and 95$^{th}$ (bottom) percentiles of the spherical radii for all (black), isolated haloes (green), haloes in pairs (yellow), main branch systems (red), and building blocks (blue). The majority of the stellar populations of proto-Milky Ways are contained within $\sim40$ kpc, with $50\%$ of the mass contained within $\sim5-10$ kpc from the host's centre at present day. In each panel, the label ``pairs'' corresponds to the systems studied in \textit{ELVIS}, and in a Local Group-like environment at $z=0$.}
\label{fig_r50r95}
\end{figure}

\begin{figure}
\centering
\includegraphics[width=\columnwidth]{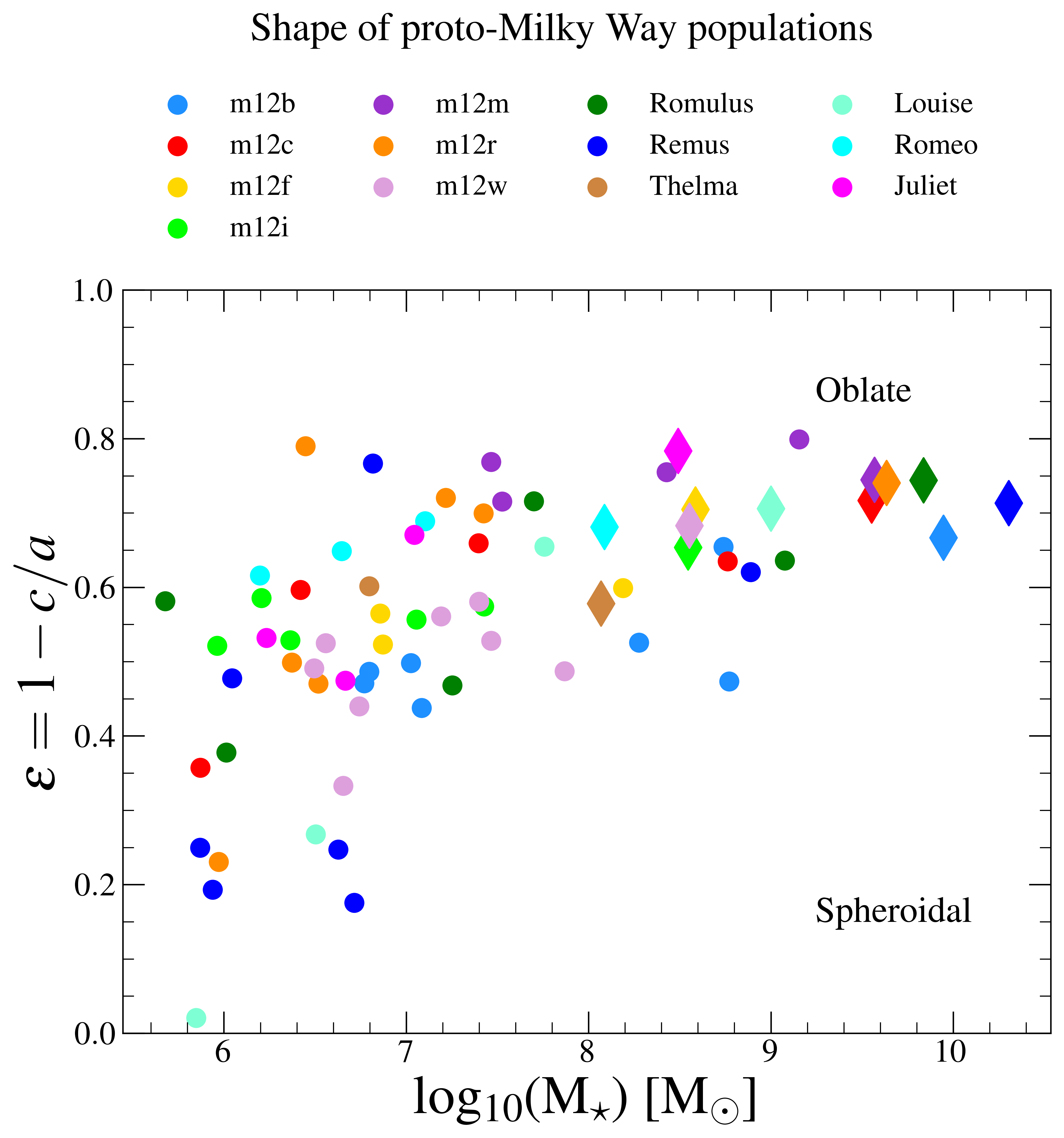}
\caption{Ellipticity (i.e., the ratio of 95$^{th}$ percentiles between cylindrical height and radius ---i.e., the aspect ratio--- subtracted from unity) as a function of the stellar mass for main branch debris (diamonds) and building block events (circles). More massive systems tend to adopt a more oblate shape.}
\label{shape_mass}
\end{figure}

As is to be expected, there is a range of spatial distributions for all the main branch and building block events. However, the majority of the mass from these systems is predominantly contained within $\sim$10 kpc from the centre of the host halo. This is especially the case when examining either main branch progenitors (dashed lines) and/or more massive building block debris (darker solid lines). For all Milky Way-mass galaxies the main branch system contains $\sim$50$\%$ of its mass within $\sim$10 kpc from the host's centre, and $\sim$95$\%$ within $\sim30-40$ kpc. For the case of the building blocks, the most massive events (M$_{\star}\gtrsim5\times10^{8}$ M$_{\odot}$) typically contain $\sim$50$\%$ of their mass within $\sim5-20$ kpc, and $\sim$95$\%$ of their mass within $\sim20-40$ kpc, in a similar fashion to the main branch population. This is especially the case for m12f, m12m, and Romulus, where the main branch progenitor follows a very similar spatial profile to the most massive building block in that system, making these almost indistinguishable spatially. For the lower mass building blocks (M$_{\star}\lesssim5\times10^{6}$ M$_{\odot}$), we see a wide range of profiles, ranging from $\sim10-50$ kpc for 50$\%$ of their enclosed mass and $\sim20-100$ kpc for 95$\%$ of their enclosed mass. This result suggests that the innermost regions of the Milky Way-mass galaxies is where you would expect to find the oldest populations in the Galaxy (\citealp{Badry2018,Fragkoudi2020}) that comprise the proto-Milky Way, arising from both a main branch progenitor and (possibly) the most massive building blocks (\citealp{Horta2021,Rix2022}). Due to the strong spatial overlap between main branch systems and massive building blocks, it would be difficult to distinguish these based on spatial distribution information alone.

We find a small difference in the spatial distribution of the debris of an event that occurs in a halo that is isolated versus a halo that is in a Local Group environment, whereby the latter tend to present proto-Milky Way populations more concentrated toward the host's centre. These results can be more easily digestible in Fig~\ref{fig_r50r95}, where we show histograms of $r_{50}$ and $r_{95}$ (i.e., the 50$^{th}$ and 95$^{th}$ percentiles of the spherical radii for each system), following the colour coding from Fig~\ref{fig_agemass}. 

 Fig~\ref{shape_mass} provides a summary of morphology by showing the aspect ratio ($c/a$) subtracted from unity, as a function of their stellar mass. As in previous figures, the main branch populations are shown as diamonds and building block systems as circles. Given that the 95$^{th}$ percentiles represent almost the extent of the distribution, we use these values to define two axes: a semi-major one ($a=R_{95}$) and a semi-minor one ($c=z_{95}$), both in units of kpc. We define these quantities so that we can measure the aspect ratio ($c/a$) and in turn their ellipticity, defined as $\varepsilon$ = 1 -- $c/a$. Given this definition, a value of $\varepsilon$ = 1 demarks a perfect circle, and an $\varepsilon$ value closer to 0 corresponds to a more squashed ellipse. We note that these aspect ratios are defined with respect to the principal axis of the host at $z=$0.

We find that the there is a dependence on $\varepsilon$ with stellar mass, whereby more massive systems tend to adopt a more oblate distribution at present day when compared to their lower mass counterparts, which adopt a wider range of morphologies. As we saw in Section~\ref{sec_massage}, the main branch debris are typically the higher mass events. This leads to conclude that main branch progenitors adopt a more oblate morphology when compared to the lower mass building block counterparts. However, we find that the more massive building block debris also adopt an oblate distribution, following the main branch progenitors.

\subsection{Kinematic properties}
\label{sec_kinematics}
In this section, we set out to examine the kinematic properties of populations comprising proto-Milky Way systems. Specifically, we aim to answer the following questions: \textit{How rotationally supported are the main branch and building block systems that form the proto-Milky Way at present day?} \textit{Are proto-Milky Way populations rotating on prograde or retrograde orbits?}

To quantify the amount of rotational support in each system, we compute the ratio of the rotational velocity over the total velocity for all star particles in a given population, $\kappa_{\mathrm{rot}}$/$\kappa_{\mathrm{tot}}$. These two velocity quantities are defined in the coordinate system centred on the host Milky Way-mass galaxy's disc at $z=0$, and are computed by finding the centre of the Milky Way-mass host using its baryonic particles at that redshift. This can be estimated by taking the ratio of the kinetic energy in the rotational direction over the total kinetic energy \citep{McCarthy2012}, such that:
\begin{equation}
    \kappa_{\mathrm{rot}} = \frac{1}{N} \sum_{i}^{N}\frac{(L_{z_{i}}/R_{i})^{2}}{2}, 
\end{equation}
assuming $R$ = $\sqrt{X^2+Y^2}$ \footnote{$R$ here is the cylindrical galactocentric radius, and $X$ and $Y$ are the galactocentric cartesian coordinates.}, $L_z = |\vec{R}\times \vec{v_{\phi}}|$, and
\begin{equation}
        \kappa_{\mathrm{tot}} = v_{\mathrm{tot}}^{2}/2,
\end{equation}
where
\begin{equation}
    v_{\mathrm{tot}} = \sqrt{v_X^2+v_Y^2+v_Z^2}.
\end{equation}
We normalise this quantity by taking the ratio of this value with the amount of rotational support determined in the young (age$<4$ Gyr) disc ($d_{\mathrm{form}}<30$ kpc) of the host of each Milky Way-mass halo, $\kappa/\kappa_{\mathrm{disc}}$ (where $\kappa=\kappa_{\mathrm{rot}}$/$\kappa_{\mathrm{tot}}$). 

\begin{figure*}
\centering
\includegraphics[width=\textwidth]{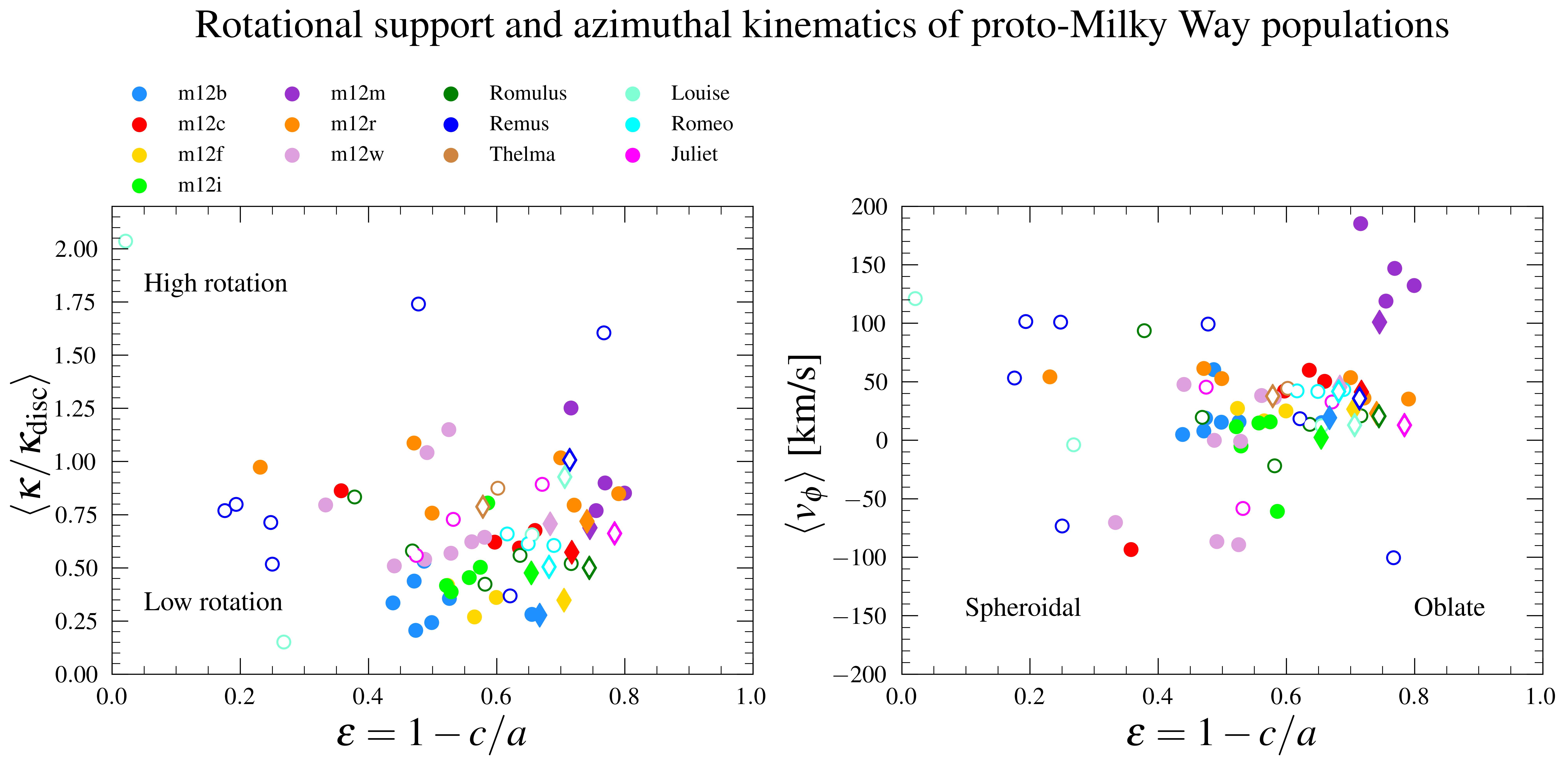}
\caption{\textit{Left}: mean of the rotational support of each proto-Milky Way population normalised by the rotational support of its host's disc ($\kappa/\kappa_{\mathrm{disc}}$) as a function of ellipticity. Rotational support is defined as the ratio of rotational energy ($\kappa_{\mathrm{rot}}$) in the direction of the host's galactic disc at present day and the total kinetic energy ($\kappa_{\mathrm{tot}}$). The disc population is comprised by star particles with age$<$4 Gyr formed in the main host, $d_{\mathrm{form}}<$30 kpc. \textit{Right}: mean azimuthal velocity in the direction of the host's disc at present day as a function of ellipticity. Systems in isolation (pairs) are shown as full (empty) markers. The majority of main branch and building block debris show low level of systematic net rotation (0.25$\lesssim$$\langle$$\kappa$/$\kappa_{\mathrm{disc}}$$\rangle$$\lesssim$0.8) when compared to their host present day disc. However, this rotational support is not zero. In fact, the majority of proto-Milky Way populations show prograde motion (0$\lesssim$$\langle v_{\phi}$$\rangle\lesssim50$ km/s), that can reach up to $\langle v_{\phi}$$\rangle\sim100-150$ km/s (m12i).}
\label{fig_vratio}
\end{figure*}

Fig~\ref{fig_vratio} shows the mean value of $\kappa$/$\kappa_{\mathrm{disc}}$ for all the main branch (diamonds) and building block (circles) systems as a function of their mean ellipticity. Milky Way-mass haloes in isolation (pairs) are shown as full (empty) markers. A value of $\kappa$/$\kappa_{\mathrm{disc}}=1$ signifies that a system is as rotationally supported as the young disc in that simulation. A value of $\kappa$/$\kappa_{\mathrm{disc}}>1$ implies that it is more rotationally supported, and a value of $\kappa$/$\kappa_{\mathrm{disc}}<1$ that it is less rotationally supported. In right panel, we plot the mean azimuthal velocity of every proto-Milky Way population. 

We find that the majority of the debris from both the main branch and building block systems have a mean $\kappa$/$\kappa_{\mathrm{disc}}$ value between 0.25 and 0.8, indicating not a strong amount of rotational support in the system when compared to the host's young disc. However, their rotational support is not 0, indicating that particles are not on purely radial orbits. In fact, their mean azimuthal velocities show that the majority of these systems have predominantly prograde orbits ($\langle v_{\phi}$$\rangle>$0 km/s). This means that they are rotating in the same direction as the disc, although at a much slower pace. Our results suggest that all these systems, either defined as main branch progenitor or building block, show weak but systematic net rotation in the plane of the host's disc at $z=0$ despite many of them appearing flattened/oblate in morphology (see Section~\ref{sec_spatial}). We find no dependence of the mean $\kappa$/$\kappa_{\mathrm{disc}}$ value of proto-Milky Way populations with environment (see the left panel of Fig~\ref{fig_vratio}).

It is important to note that recent studies have reported that Milky Way-mass galaxies in FIRE experience different phases of growth (\citealp{Yu2021,Gurvich2023,Hopkins2023,McCluskey2023,Semenov2023,Yu2023}), which can be characterised based on differences in their stellar kinematics and amount of rotational support. \citet{Yu2023} find that the orbits of star particles formed in the main branch spheroid at early times are more radial than those formed at later times in a more settled disc (see also \citealp{Gurvich2023} and \citealp{McCluskey2023}). This result is related to our finding, and suggests that populations formed at early times are likely to not be on as rotationally supported orbits when compared to populations formed later in settled discs.

\subsection{Chemical compositions}
\label{sec_chemistry}
In this Section, we examine what is likely to be the most pristine observable tracer of stellar populations: their chemical compositions. While FIRE tracks eleven different elemental species, they trace three nucleosynthetic channels: contributions from core-collapse supernovae, (CCSN), SN type Ia, and stellar winds. We set out to explore the distribution of the debris of these events in the chemical plane tracing the contribution of CCSN and SNIa, choosing Mg as our $\alpha$ tracer. The main question we aim to tackle in this Section is: \textit{what differences in the chemical compositions (and thus, the star formation histories) can we expect from main branch progenitors when compared to their building block counterparts?}

\begin{figure*}
\centering
\includegraphics[width=1\textwidth]{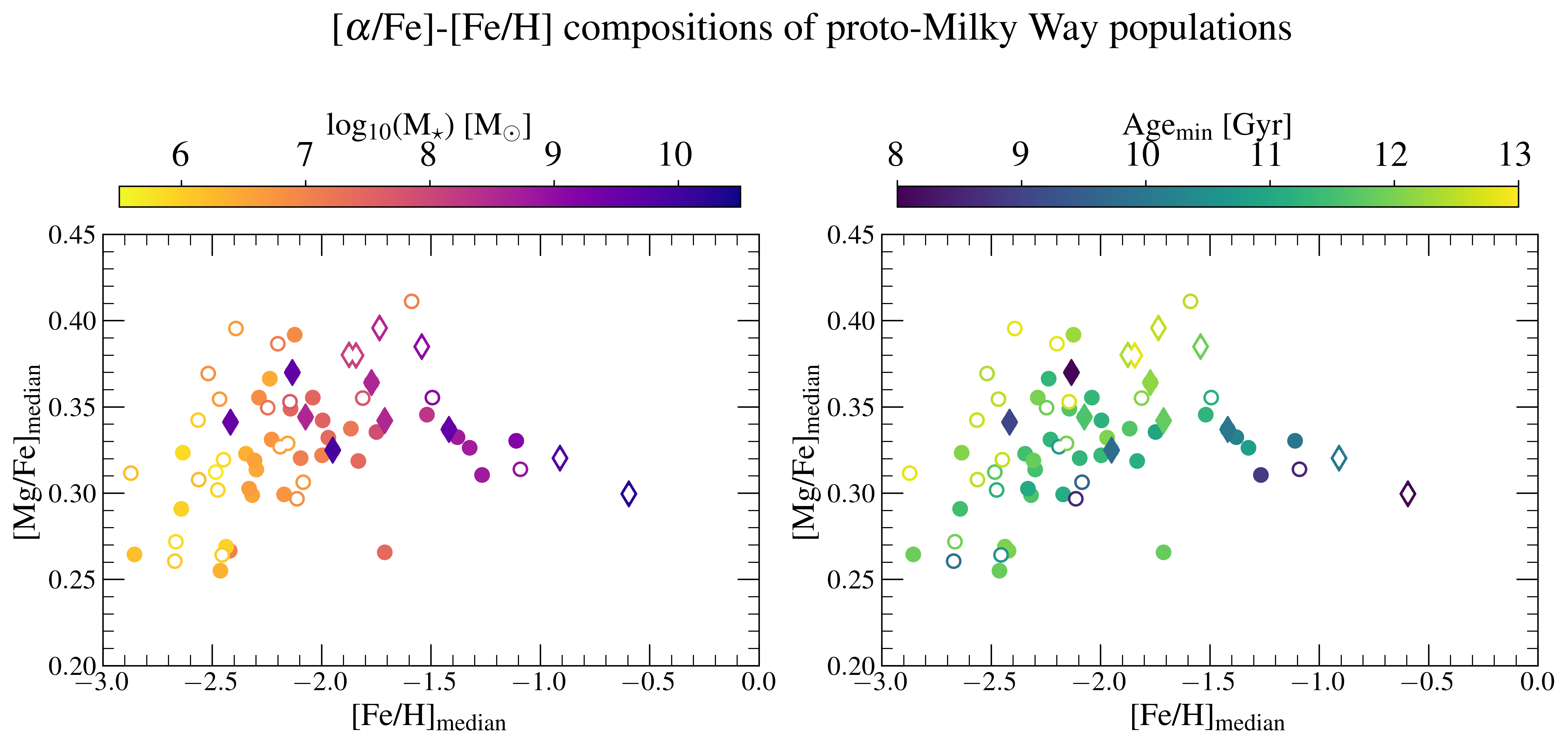}
\caption{Median [Mg/Fe] and [Fe/H] values for main branch (diamonds) and building block (circles) populations for isolated Milky Way-mass haloes (filled markers) and those in pairs (empty markers), colour coded by stellar mass (left) and minimum star particle age (right). The average 1-$\sigma$ is 0.33 dex for [Fe/H], and 0.08 dex for [Mg/Fe]. On average, main branch progenitors and the most massive building blocks present more enriched chemical compositions when compared to lower mass building blocks; this is a result of the galaxy mass-metallicity relation.}
\label{alpha-fe}
\end{figure*}

Fig~\ref{alpha-fe} shows the median [Mg/Fe] and [Fe/H] values for building blocks (circles) and main branch debris (diamonds), where each system is colour coded by their mean stellar mass (left) and minimum star particle age (right). Empty markers correspond to systems in Local Group-like environments (i.e., the \textit{ELVIS} suite) and filled markers correspond to events in isolated haloes (i.e., the \textit{Latte} suite).

Fig~\ref{alpha-fe} demonstrates that, on average, the main branches are more metal-rich than the building blocks. Here, main branch progenitors either have a higher [Mg/Fe] value at fixed [Fe/H], or higher [Fe/H] overall. However, we find that this is also the case for the most massive (and youngest) building blocks. This is a natural consequence of the mass-metallicity relation (\citealp[e.g.,][]{Kirby2013,Kirby2020}). More massive systems are able to form stars for longer, thus enriching their interstellar medium with more metals and evolving chemically faster than lower mass systems that quench star formation earlier. This fact could be leveraged to disentangle the lower mass building blocks from their main branches using their abundances (e.g., employing methods presented in \citealt{Cunningham2022}, \citealt{Horta2023_abun}, \citealt{Deason2023}, for example). 

\begin{figure}
\centering
\includegraphics[width=\columnwidth]{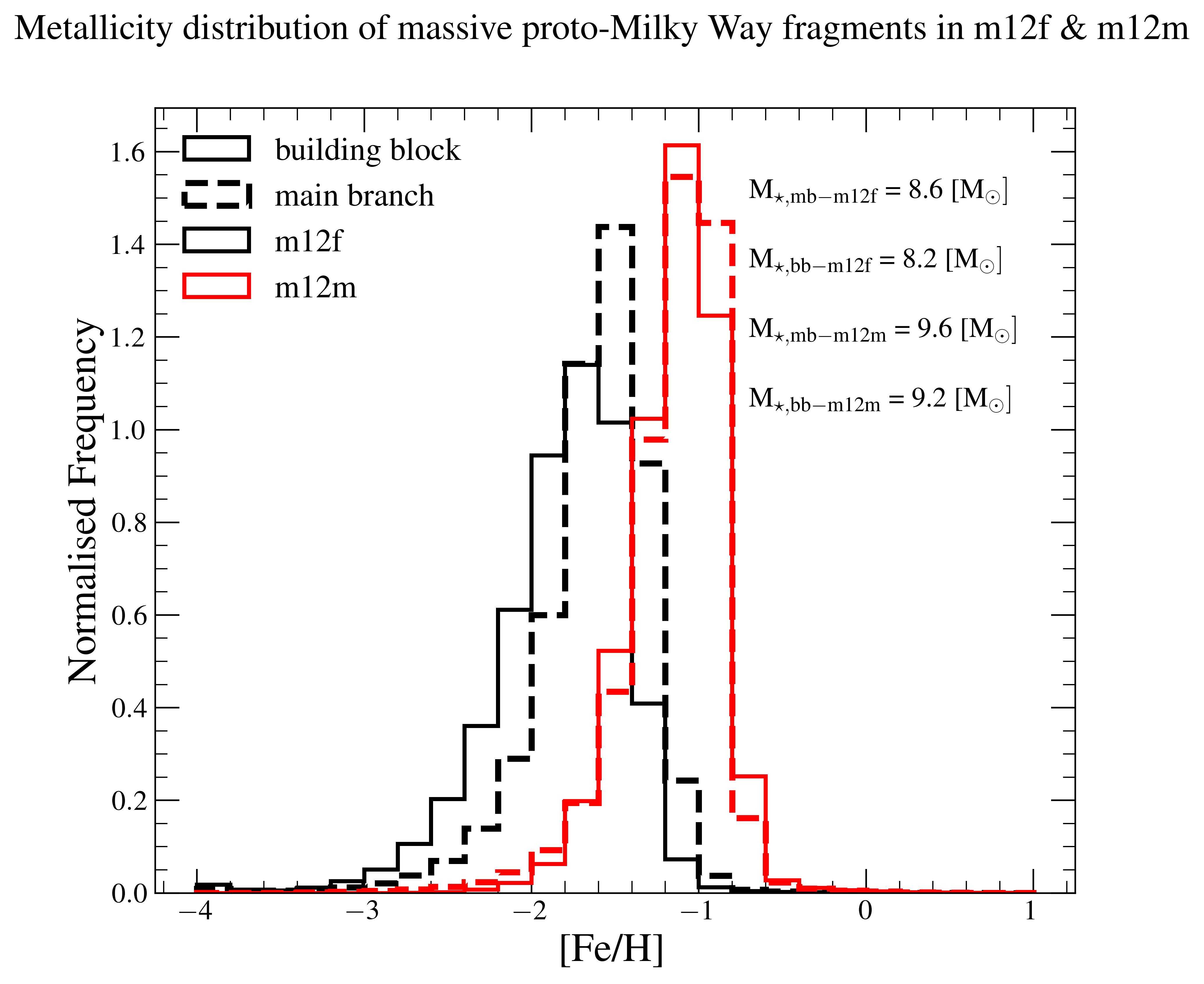}
\caption{Histograms of the metallicity distribution functions (MDF) of main branch progenitors (dashed) and the most massive building blocks (solid) events in the m12f (black) and m12m (red) simulations. The MDF profiles are qualitatively the same for these populations. This result highlights the difficulty of distinguishing main branch from massive building block populations based on their MDFs.}
\label{mdf}
\end{figure}

However, some of the more massive building blocks have chemical compositions consistent with the main branches. As a result, it could be challenging to disentangle populations from the main branch from populations from the most massive building block when a proto-Milky Way has two significant contributors (which is the case for roughly $\sim$40$\%$ of our Milky Way-mass galaxy sample). We investigate this further in Fig~\ref{mdf}, in which we compare the metallicity distribution functions (MDFs) for the main branch (dashed) and most massive building block (solid) in both m12f (black) and m12m (red). In the case of m12f, the mean of the main branch MDF is $\sim0.1-0.2$ dex higher compared to the most massive building block. Conversely, the main branch in m12m is identical in [Fe/H] when compared to its massive building block counterpart. These results show that, quantitatively, it would be extremely difficult to distinguish massive building blocks from their main branch system counterparts using their MDFs. More detailed chemical abundance information from elemental species synthesised in more exotic nucleosynthetic channels, which we have not been able to examine in this work (but see \citealp{Horta2023_abun}), may hold the clues to disentangling these dominant proto-Milky Way fragments.

When comparing the debris from host haloes that evolve in different environments, we find that there are possible subtle differences. Specifically, we find that the main branch and building block events in host haloes simulated in Local Group-like environments (pairs) present on average slightly higher median [Mg/Fe] and/or [Fe/H] values when compared to isolated hosts. This result is consistent with our finding that systems in Local Group-like environments evolve either faster and/or earlier, when compared to systems in isolated hosts \citep{Santistevan2020}. However, within the spread of the distributions (i.e., the uncertainties in Fig~\ref{alpha-fe}), which is on average $\sim0.05-0.1$ dex for [Mg/Fe] and $\sim0.3$ dex for [Fe/H], we note that this result is tentative.

\section{Discussion}
\label{sec_discussion}

\begin{figure*}
\centering
\includegraphics[width=1\textwidth]{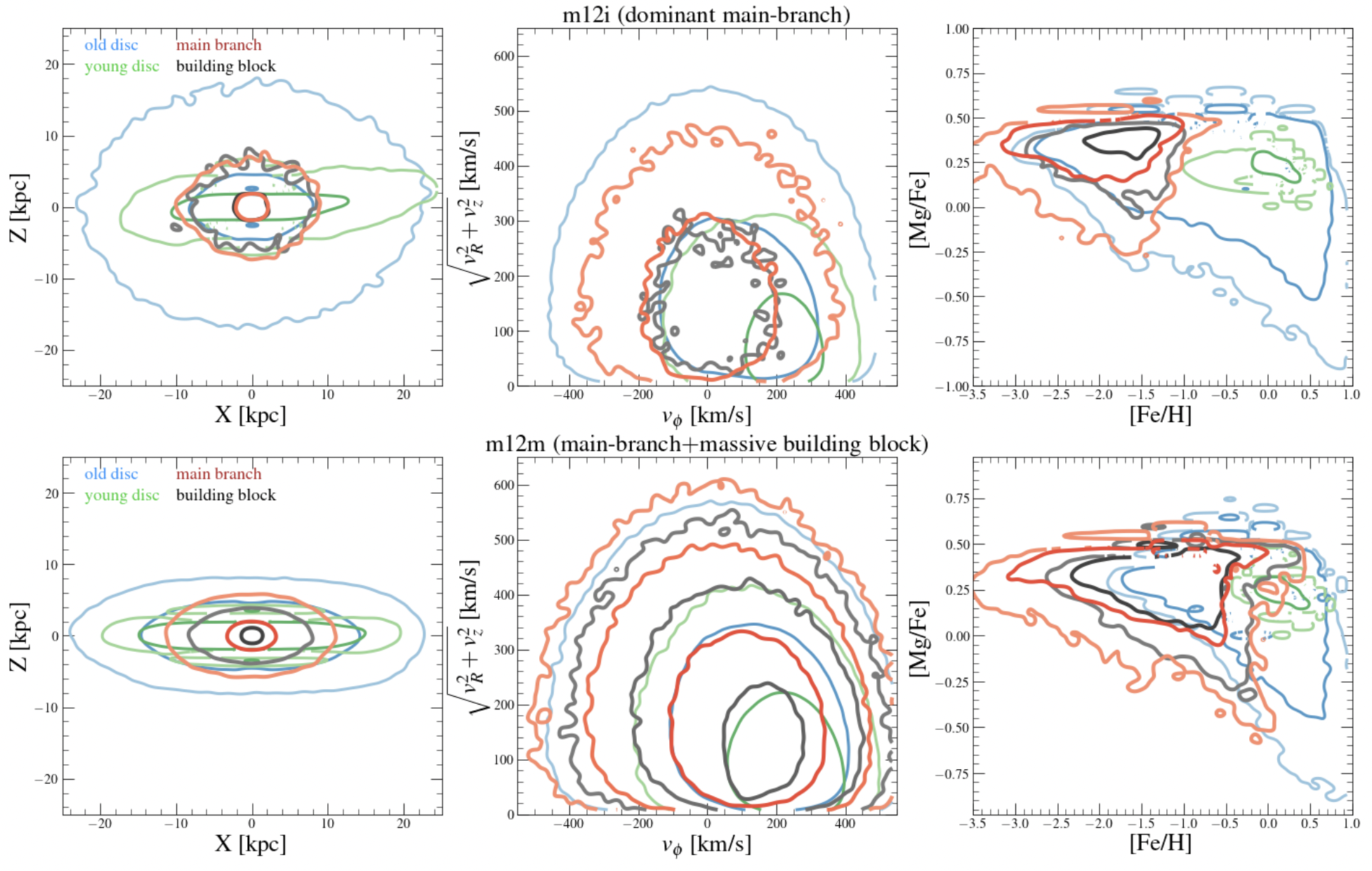}
\caption{\textbf{\textit{Left}}: Density level contours of the X-Z positions (i.e., edge-on projections) at $z=0$ for star particles belonging to the main branch system (red), most massive building block (black), the old disc (blue), and the young disc (green) in m12i. Here, the old(young) disc is defined as star particles formed in the main host halo between 4$<$$\tau$[Gyr]$<$$t_{\mathrm{MR_{3:1}}}$(4$<$$\tau$[Gyr]). \textbf{\textit{Middle}}: Same as left, but now in the cylindrical Toomre diagram. \textbf{\textit{Right}}: The same as left, but now in the [Mg/Fe]-[Fe/H] plane. m12i and m12m differ in the fact that the former proto-Milky Way is comprised primarily by one main branch system ($\sim$90$\%$ of the stellar mass) and m12m is comprised by two systems, a main branch system ($\sim$65$\%$ of M$_{\star}$) and a massive building block ($\sim$35$\%$ of M$_{\star}$). The kinematic distributions of main branch, building block, and old disc are qualitatively similar and are different from the young disc. This is more pronounced in m12i than m12m. The [Mg/Fe]-[Fe/H] compositions of the main branch and building block system are also extremely similar, but are different from the young disc. The main branch and building block overlap with the metal-poor old-disc; this overlap is bigger for m12m than or m12i.}
\label{comp}
\end{figure*}

\subsection{Can we distinguish if a proto-Milky way formed from one or two dominant systems?}
\label{sec_2v1}

Fig~\ref{comp} shows the X-Z positions, Toomre diagram distribution, and [Mg/Fe]-[Fe/H] density contour distributions for the main branch (red), the most massive building block (black), the old disc (blue), and young disc (green) in the m12i and m12m simulations. Disc stars are selected to be formed within 30 kpc of the main host after $t_{\mathrm{MR_{3:1}}}$, where the old (young) disc is older (younger) than 4 Gyr. We choose to compare m12i with m12m, as m12i is a clear case where the proto-Milky Way formed from one dominant main branch system (M$_{\star}=$ 3.51$\times$10$^{8}$M$_{\odot}$), and m12m is a clear example of a proto-Milky Way formed from two dominant systems (a main branch  of mass M$_{\star}=$ 3.69$\times$10$^{9}$M$_{\odot}$ and a massive building block of mass M$_{\star}=$ 1.43$\times$10$^{9}$M$_{\odot}$). As can be seen in Fig~\ref{comp}, there are subtle differences in the spatial distribution (at $z=0$) for all populations examined between m12i and m12m. More specifically, m12i presents a clear thin young disc population in addition to a spherical old disc, main branch, and building block (of mass M$_{\star}=$ 2.67$\times$10$^{7}$M$_{\odot}$). Conversely, m12m presents a much flatter distribution, where the main branch and massive building block systems present an oval shape, in a similar fashion to the old disc in this Milky Way-mass galaxy. 

Moreover, we find that in m12i, the main branch and building block overlap in the Toomre diagram and are largely non-rotational ($v_{\phi}$$\sim$0 km/s), and are different to the old/young disc, which show a more extended distribution rotating at higher $v_{\phi}$. Conversely, m12m reveals that its main branch and massive building block also overlap in kinematic space, but rotate at a higher tangential velocity of $v_{\phi}$$\sim$100 km/s, lagging behind the young disc by $\sim$150 km/s. In this halo, the main branch and massive building block also overlap with the old disc. The difference in the $v_{\phi}$ magnitude between the proto-Milky Way system in m12i and m12m is interesting. However, it is likely due to m12i being an outlier in terms of the rotational velocities of its metal-poor stars \citep{Santistevan2022}. All the other \textit{Latte} Milky Way-mass galaxies show a preference for prograde disc orbits for older and/or lower metallicity stars, similar to m12m (although this system also shows higher than average rotation, see the right panel of Fig~\ref{fig_vratio}). \citet{Santistevan2022} argue that metal-poor/old stars on prograde disc orbits are a consequence of major building blocks depositing stars/gas on prograde orbits, which typically set the orientation of the resulting Milky Way-mass galaxy disc at $z=0$.

\begin{figure}
\centering
\includegraphics[width=0.8\columnwidth]{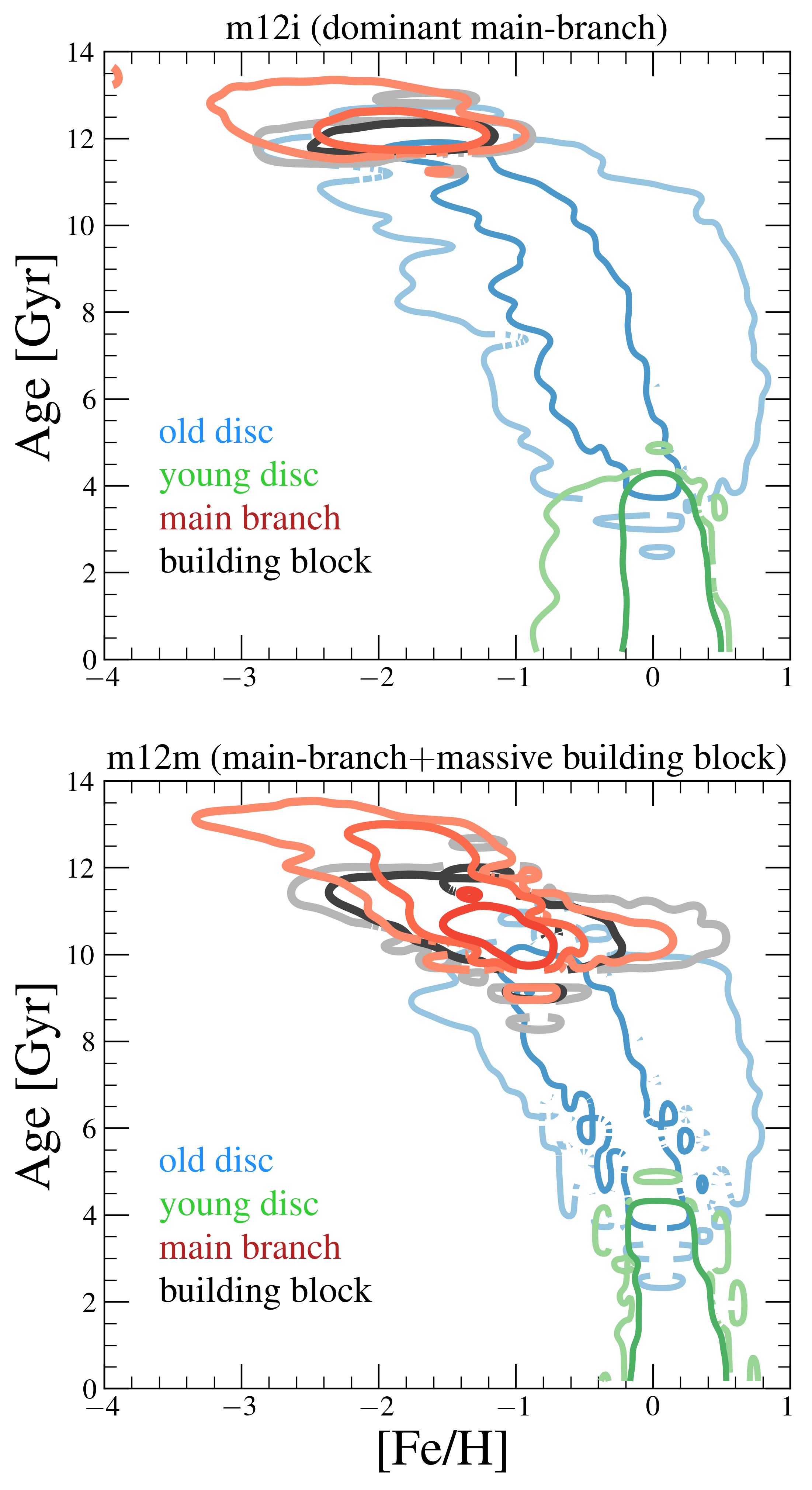}
\caption{Age metallicity relation for the same samples as shown in Fig~\ref{comp}. For this particular comparison we see that m12i, a Milky Way-mass halo with a proto-galaxy population formed primarily by one massive and dominant halo (i.e, main branch), is older than the proto-galaxy population of m12m, a Milky Way-mass halo formed from two dominant halos (i.e., a main branch and a massive building block). We argue that this $\sim2$ Gyr difference should be detectable in relative ages of metal-poor stars. Moreover, the peak in the metallicity distribution for the proto-Milky Way population in m12i is more metal-poor than the one in m12m, on the order of $\sim$1 dex.  These subtle differences are also seen for $\sim60\%$ of the parent Milky Way-mass galaxy sample (see Section~\ref{sec_2v1} for details). These results hint that age estimates and [Fe/H] information for large samples of metal-poor stars in the inner regions of the Milky Way could possibly help answer if the Galaxy formed from one dominant halo or from two.}
\label{age-metals}
\end{figure}

In terms of their [Mg/Fe]-[Fe/H] compositions, we find again that for both m12i and m12m, the main branch and building block populations overlap. This is to be expected for m12m (see Fig~\ref{alpha-fe} and Fig~\ref{mdf}), but is a surprise for m12i. Interestingly, the [Mg/Fe]-[Fe/H] compositions of these systems are clearly different from that of the old/young discs in m12i, which present a much higher [Fe/H] distribution. However, we note that there is an overlap between the metal-rich sequence of the main branch/building block and the metal-poor tail of the old disc. Conversely, for m12m this overlap between the main branch and massive building block is more pronounced. This highlights that the dominant proto-Milky Way populations are likely to overlap with the metal-poor/old disc (\citealp{Horta2021,Conroy2021,Mardini2022,Horta2023}).

In a similar vein, another interesting diagram to investigate relates to the time difference between proto-Milky Way systems that form from one main system versus two main haloes. To investigate this point further, in Fig~\ref{age-metals} we show the main branch, (massive) building block, and old/young disc in the age-metallicity plane. Here the data is displayed in the same way as in Fig~\ref{comp}. We find that for this particular comparison, a Milky Way-mass halo with a proto-galaxy population formed primarily by one massive and dominant main branch halo, m12i (but also Thelma, Louise, Romeo, and Juliet, see Fig~\ref{fig_agemass}) is older than the proto-galaxy population of a Milky Way-mass halo formed from one dominant main branch halo and a massive building block halo, m12m (but also m12c and Romulus). This difference is on the order of $\sim$2 Gyr. Furthermore, the peak in the metallicity distribution for the proto-Milky Way population in m12i is more metal-poor than the one in m12m, on the order of $\sim$1 dex. Although this comparison has only been shown for two haloes, we find that it is qualitatively satisfied for 8 of the 13 Milky Way-mass galaxies. We argue that this result is potentially very important, and suggests that age differences in metal-poor stars in the inner regions of Milky Way-mass haloes could possibly help decipher if proto-Milky Way systems formed from one or two main haloes. However, we do note that for m12b, m12f, m12r, m12w, and Remus, this is not the case.

In summary, we argue that disentangling if a proto-Milky Way system formed from one or two dominant systems is going to be challenging, as their obervable properties are going to be similar. However, we suggest that with large and complete samples of metal-poor stars, that include elemental abundances synthesised in nucleosynthetic sites not explored in this work, as well as accurate age estimates, may hold the clues to answering this question.

\subsection{Hunting for the proto-galaxy in the Milky Way}

Given all our findings, we now provide a list of ideas/pointers we suggest one to follow if aiming to identify the majority of the populations belonging to the proto-Milky Way:\\

    \textit{i}) \textit{Mass and age}: we suggest that the best possible way to find the proto-Galaxy in the Milky Way would be to identify stellar populations older than $\tau$$\gtrsim$8 Gyr, that amount to a mass smaller than M$_{\star}\lesssim10^{9}--10^{10}~\mathrm{M}_{\odot}$.\\
    
    \textit{ii}) \textit{Spatial distribution}: this stellar population would primarily be concentrated towards the inner regions of the Galaxy (95$\%$ of the mass within $r \sim30-40$ kpc), with 50$\%$ of the mass contained within $\sim5-10$ kpc. It would also likely have an oblate profile ($\varepsilon$$\sim$0.7).\\
    
    \textit{iii}) \textit{Kinematics}: the proto-Milky Way would not be a strongly rotating population (w.r.t the present day Galactic disc, 0.25$\lesssim$$\langle$$\kappa$/$\kappa_{\mathrm{disc}}$$\rangle$ $\lesssim$0.8). However, it is still likely to host prograde orbits and manifest moderate tangential velocity values (0$\lesssim$$v_{\phi}$$\lesssim$60 km/s, see Fig~\ref{fig_vratio}), matching that of an old and metal-poor disc (Fig~\ref{comp}). \\
    
    \textit{iv}) \textit{Chemical compositions}: the proto-Milky Way could have a wide range of [Fe/H] and [$\alpha$/Fe] values. We find that the majority of the stars comprising the proto-Milky Way (either from one main branch system or a main branch and a massive building block) are likely to overlap with the metal-weak old disc in [$\alpha$/Fe] and [Fe/H]. However, additional elemental abundance ratios, which have not been explored in this work, may provide additional separation between proto-Galaxy fragments and other co-spatial populations.\\

Hunts for the proto-Galaxy and/or the building blocks that formed it have been performed recently in observational studies (\citealp[][]{Horta2021,Belokurov2022,Rix2022}) using chemical-kinematic information. These results are shedding light on the earliest stages of formation of the Milky Way. However, open questions still remain, such as: \textit{how many systems comprise the proto-Milky Way?} \textit{when did the proto-Milky Way emerge?} \textit{what caused the proto-Galaxy to emerge and morph into the metal-poor disc?} We believe that this study has provided some intuition on how to answer some of these questions, as we have been able to assess: a) what proto-Milky Way populations are made of; b) how massive and/or old proto-Milky Way systems are; c) how and if different proto-Milky Way fragments can be distinguishable in chemical-kinematic samples.

The advent of the SDSS-V project Milky Way Mapper (\citealp{kollmeier2017sdssv}) will deliver precise elemental abundance ratios for over $\sim$50,000 stars with [Fe/H]$<-$1, that when complemented with the \textit{Gaia} mission, and other upcoming massive spectroscopic surveys (e.g., WEAVE: \citealp{Dalton2012}; 4MOST: \citealp{deJong2019}), will provide an unprecedented amount of chemical-kinematic information for metal-poor stars in the innermost Galaxy. Along those lines, recent work with the \textit{Gaia} XP spectra are delivering a colossal amount of metallicities for stars in the inner regions of the Galaxy (\citealp{Andrae2023}), which are also helping resolve populations in this region \citep{Rix2022}. In a similar vein, detailed spectroscopic follow up (e.g., PRISTINE: \citealt{Starkenburg2017}) of metal-poor stars in the innermost regions of the Galaxy can also help disentangle the earliest stages of formation of the Milky Way (\citealp{Arentsen2020,Arentsen2020b,Lucey2020,Sestito2023}).

Given our results, we suggest that with these data it should be possible to pick out the massive contributors to proto-Milky Way systems from the smaller building blocks. We are only scratching the surface, but we are on the right path to finding the Milky Way's heart and understanding its early assembly history.

\section{Limitations and further considerations}
\label{sec_limitations}

Before listing our conclusions, it is imperative we discuss the limitations of our work. Below we provide a summary of the limitations and considerations for the interpretations of our findings:

\begin{itemize}
    \item The new aspect of this work has been the ability to track the systems that comprise proto-Milky Way populations. However, in doing so, we have decided to track only those luminous halos that are resolvable given the resolution limits of FIRE-2. Specifically, we have tracked the populations from halos with 150 star particles or more, leading to a minimum stellar mass of $\sim$1$\times$10$^{6}$ M$_{\odot}$ for the \textit{Latte} suite, and $\sim$5$\times$10$^{5}$ M$_{\odot}$ in the \textit{ELVIS} suite\footnote{The small difference in the resolution between the two suites is small enough to not have any impact on our results, as we are tracking haloes with 150 star particles or more.}. Albeit our hands being tied with the ability to resolve haloes, we have been cautious in our choices to be able to track as many resolvable systems as possible in these simulations. However, it is likely that in reality there are more luminous halos of lower masses that contribute to the build-up of proto-Milky Way populations. We argue that the contribution in mass of these lower mass building blocks is small, but cannot rule out the possibility of it being non-negligible (see \citealp{Santistevan2020} and \citealt{Ghandi2023} for an accounting of lower mass building blocks). Furthermore, the kinematic properties of the lower mass haloes may not be fully resolved with 150 star particles. Although this only affects a small fraction of our sample, it is another limitation to keep in mind. Studies focusing on tackling the assembly history of proto-Milky Way populations, and the properties of their constituent main branch and building block systems, may need to take this issue into account when comparing to our findings.
    \item A key property we have defined in this work has been $t_{\mathrm{MR_{3:1}}}$, the time at which a proto-Milky Way emerges. Although well reasoned (see Section~\ref{sec_proto_bbs} for details), this choice is arbitrary, and has ramifications on all the present day properties of the proto-Milky Way populations, as well as the number of building block systems each proto-Milky Way inherits. It could have also been equally as valid to define this time in another well motivated way (for example, the time in which the host halo peaks in star formation). This is beyond the scope of this work, but could be an interesting way of expanding on these results.
    \item Here we have only studied the properties of the field stellar components of proto-Milky Way populations. However, galaxies across the Cosmos contain globular clusters (GCs). In the Milky Way, it has been shown that the disruption of GCs contribute a significant amount to the total stellar halo mass budget (\citealp[][]{Martell2017,Schiavon2017_nrich,Koch2019,Horta2021_nrich, Belokurov2023}). This would imply that in addition to the field components comprising proto-Milky Way systems, one must also take into account the contribution from disrupted/evaporated GCs. For this work, this limitation is due to FIRE-2 not including prescriptions for the formation and evolution of GCs. However, we argue that it is an important point that needs to be considered for observational and potential future simulation work.
    \item When comparing the observable properties of main branch and building block events, we have primarily only examined qualitatively the mean/median values. It would be interesting to investigate how the average and full distribution of the spatial, kinematic, and chemical properties of these systems compare quantitatively (following methods and tools that already exist; e.g., \citealp{Cunningham2022,Horta2023_abun}). 
\end{itemize}

\section{Conclusions}
\label{sec_conclusions}

At the earliest stages of formation, galaxies experience rapid and chaotic growth, either by coalescence of low mass galaxies/clumps and/or filamentary supply of gas. In the FIRE-2 simulations, bursty stellar feedback that repeatedly blows apart the ISM at early times also appears to be critical to set the properties of early stellar populations (\citealp[e.g.,][]{Yu2021,Gurvich2023,Hopkins2023}). The remains of the stars born during this phase constitute the proto-galaxy, and should retain the clues to understanding galaxy formation at these earliest stages. In this work, we have searched for the fragments (namely, the main branch and building blocks) that constitute the proto-galaxy in thirteen Milky Way-mass haloes from the FIRE-2 simulations. We then examined their observable properties at present day, with the aim of answering the following questions:

\begin{itemize}
    \item \textbf{What constitutes a proto-Milky Way?} We find that proto-Milky Way populations are made of either one ($\sim60\%$) or two ($\sim40\%$) dominant systems of similar mass to the LMC (i.e., M$_{\star}\sim1\times10^{9}$M$_{\odot}$) and $\sim3-5$ other lower-mass building blocks with an average stellar mass of M$_{\star}\sim4\times10^{7}$M$_{\odot}$ (see Fig~\ref{fig_summary_mass}, Fig~\ref{fig_agemass}, and Fig~\ref{mass-ratios}). The case of two clear dominant systems comprising the proto-Milky Way is especially clear in m12f and m12m. However, the number of building blocks we find in this study that comprise a proto-Milky Way is grounded by our choice to only track systems with 150 star particles or more. 
    \item \textbf{When does a proto-Milky Way form?} Given our assumptions (Section~\ref{sec_proto_bbs}) we find that on average, proto-Milky Way populations are old (see Table~\ref{tab1}). Their minimum age can range from $\mathrm{Age_{min}} \gtrsim 8.05$ Gyr ($z=1.02$) to $\mathrm{Age_{min}} \lesssim 12.9$ Gyr ($z=6.08$). We find that overall the proto-Milky Way systems can be grouped into three main camps: an early forming group, an intermediate forming group, a late forming one.
    \item \textbf{Does environment play a role?} The noticeable differences found between galaxies in different environments are the times in which proto-Milky Way systems assemble, and the sizes of the average building block. Table~\ref{tab1} shows that on average, systems in pairs assemble earlier and from smaller mass systems than proto-Milky Way populations in isolation, in line with results from \citet{Santistevan2020}. We also find that pairs tend to contain 50$\%$ of their stellar populations closer to the host's centre when compared to isolated systems (Fig~\ref{fig_r50r95}), and that pairs tend to present slightly more enriched median [$\alpha$/Fe] and/or [Fe/H] composition values.
    \item \textbf{Where are the debris of proto-Milky Way systems spatially contained?} The dominant components of the proto-Milky Way (namely, the main branch population and the most massive building blocks) contain 50$\%$ of their stellar mass within $r\lesssim5-10$ kpc, and 95$\%$ within $r\lesssim40-60$ kpc (see Fig~\ref{radii} and Fig~\ref{fig_r50r95}). 
    \item \textbf{What shape do debris from the proto-Milky Way have?} Although different main branch/building blocks adopt a range of morphologies, the dominant components of proto-Milky Way systems adopt an oblate shape ($\varepsilon\sim0.7$, see Fig~\ref{shape_mass}).
    \item \textbf{What kinematics do the debris of proto-Milky Way systems have?} All the fragments of proto-Milky Way systems show low level of systematic net rotation with respect to the present day disc of the host galaxy (i.e., $\langle\kappa$/$\kappa_{\mathrm{disc}}\rangle\lesssim0.8$), but are also not purely isotropic (Fig~\ref{fig_vratio}). The majority of these stellar populations rotate on prograde orbits, and can reach average azimuthal velocities of up to $v_{\phi}\sim100-150$ km/s.
    \item \textbf{What are the chemical compositions of proto-Milky Ways?} The main branch and building block components of proto-Milky Way systems can present a wide range of [Fe/H] and [$\alpha$/Fe] values, and can adopt high [$\alpha$/Fe] values (Fig~\ref{alpha-fe}). This would make it difficult to distinguish the dominant proto-Milky Way populations from populations formed later in the disc using the average value of these abundances alone (see Fig~\ref{comp}). However, disentangling the massive fragments from the lower mass ones is possible with [$\alpha$/Fe] and [Fe/H] compositions; similarly, additional element abundances may also help distinguish the larger mass fragments (e.g., \citealp{Horta2023_abun}).
    \item \textbf{Can we distinguish the fragments that build up proto-Milky Way systems?} On the whole, it is possible to separate the dominant components of proto-Milky Way systems (i.e., the main branch) from the non-dominant (low mass building blocks) components using chemical-kinematic samples. However, distinguishing stars formed in the main branch progenitor from the most massive building blocks will likely be difficult owing to the big overlap in all chemical and kinematic planes shown in this work. However, studies focusing on other chemical abundance measurements and/or studies with larger samples may be able to disentangle the dominant proto-Milky Way fragments based on chemical abundance measurements.
    \item \textbf{How do we hunt for the proto-Milky Way?} We suggest that metal-poor stars confined to the inner Galaxy are likely to be the easiest to find, as they are the debris from the more dominant fragments that formed the proto-Milky Way. These stars should have prograde orbits that are not strongly rotating with the Milky Way's disc, but are also not purely isotropic. They should present, on average, more enriched chemical abundance ratios when compared to lower mass building blocks and later accreted satellite galaxies, and are likely to overlap in chemical space with the metal-poor tail of the old disc. Thus, we suggest that additional chemical abundance information, likely probing different nucleosynthetic channels may help disentangle different fragments of the proto-Milky Way.
    \item \textbf{Is the proto-Milky Way formed from one or two dominant haloes?} It may be possible to answer this question by examining relative age differences between metal-poor stars in the central regions of the Galaxy. From our comparison of m12i and m12m in Fig~\ref{age-metals}, proto-Milky Way systems formed from one dominant halo may tend to assemble earlier, and are thus older and more metal-poor, when compared to proto-Milky Way systems formed from two dominant haloes. However, we do stress that this result is from one comparison alone, and is only applicable to $\sim60\%$ of the full sample (8 of the 13 Milky Way-mass galaxies studied). 
\end{itemize}

\section*{Acknowledgements}
DH thanks Ricardo P. Schiavon, Vasily Belokurov, Hans-Walter Rix, and Stephanie Monty for helpful discussions. He also thanks Sue, Alex, and Debra for their constant support. ECC acknowledges support for this work provided by NASA through the NASA Hubble Fellowship Program grant HST-HF2-51502 awarded by the Space Telescope Science Institute, which is operated by the Association of Universities for Research in Astronomy, Inc., for NASA, under contract NAS5-26555. RES gratefully acknowledges support from NSF grant AST-2007232 and NASA grant 19-ATP19-0068. CAFG was supported by NSF through grants AST-2108230  and CAREER award AST-1652522; by NASA through grants 17-ATP17-0067 and 21-ATP21-0036; by STScI through grant HST-GO-16730.016-A; and by CXO through grant TM2-23005X. PJG received support from the Texas Advanced Computing Center (TACC) via a Frontera Computational Science Fellowship.

\section*{Data Availability}
FIRE-2 simulations are publicly available \citep{Wetzel2023} at \url{http://flathub.flatironinstitute.org/fire}.
Additional FIRE simulation data is available at \url{https://fire.northwestern.edu/data}.
A public version of the \textsc{Gizmo} code is available at \url{http://www.tapir.caltech.edu/~phopkins/Site/GIZMO.html}.


\bibliographystyle{mnras}
\bibliography{refs} 

\appendix

\bsp	
\label{lastpage}
\end{document}